\documentclass[aps, preprint]{revtex4}
\usepackage{graphics}

\begin{document}

\newcommand{\n}{\eta}
\newcommand{\nb}{\bar{\eta}}
\newcommand{\np}{\eta^{\prime}}
\newcommand{\nbp}{\bar{\eta}^{\prime}}
\newcommand{\bk}{{\bf k}}
\newcommand{\bl}{{\bf l}}
\newcommand{\bq}{{\bf q}}
\newcommand{\br}{{\bf r}}
\newcommand{\z}{z}
\newcommand{\zb}{\bar{z}}
\newcommand{\zp}{z^{\prime}}
\newcommand{\zbp}{\bar{z}^{\prime}}
\newcommand{\la}{\lambda}
\newcommand{\lab}{\bar{\lambda}}
\newcommand{\w}{\omega}
\newcommand{\e}{\epsilon}
\newcommand{\psib}{\bar{\psi}}
\newcommand{\psip}{\psi^{\prime}}
\newcommand{\psibp}{\bar{\psi}^{\prime}}
\newcommand{\phib}{\bar{\phi}}
\newcommand{\phip}{\phi^{\prime}}
\newcommand{\phibp}{\bar{\phi}^{\prime}}
\newcommand{\g}{\Gamma}
\newcommand{\s}{\sigma}

\title{Quantum Entanglement under Non-Markovian Dynamics of Two Qubits Interacting with a common Electromagnetic Field}
\author{C.~Anastopoulos$^{1}$\footnote{Corresponding author. Email
address: anastop@physics.upatras.gr},
S.~Shresta$^{2,3}$\footnote{Present address: MITRE Corporation 7515
Colshire Drive, MailStop N390 McLean, VA 22102. Email Address:
sanjiv$\_$shresta@mitre.org},  and B.~L. Hu$^{2}$\footnote{Email
address: blhu@umd.edu}} \affiliation{$^1$Department of Physics,
University of Patras, 26500 Patras, Greece, \\$^2$Department of
Physics,
University of Maryland, College Park, Maryland 20742-4111 \\
$^3$NIST, Atomic Physics Division, Gaithersburg, MD 20899-8423}

\date{January 31, 2007}

\begin{abstract}
We study the non-equilibrium dynamics of a pair of qubits made of
two-level atoms separated in space with distance $r$ and interacting
with one common electromagnetic field 
but not directly with each other. Our calculation makes a weak
coupling assumption 
but no Born or Markov approximation. We derived a non-Markovian
master equation for the evolution of the reduced density matrix of
the two-qubit system after integrating out the electromagnetic field
modes. It contains a Markovian part with a Lindblad type operator
and a nonMarkovian contribution, the physics of which is the main
focus of this study. We use the concurrence function as a measure of
quantum entanglement between the two qubits. Two classes of states
are studied in detail: Class A is a one parameter family of states
which are the superposition of the highest energy $|I \rangle \equiv
|11 \rangle$ and lowest energy $|O \rangle \equiv |00 \rangle$
states, {\it viz}, $|A \rangle \equiv \sqrt p|I \rangle + \sqrt
{(1-p)} |O \rangle$, with $ 0 \le p \le 1 $; and Class B states $|B
\rangle$  are linear combinations of the symmetric $|+ \rangle =
\frac{1}{\sqrt 2} (|01 \rangle + |10 \rangle)$ and the antisymmetric
$|- \rangle = \frac{1}{\sqrt 2} (|01 \rangle - |10 \rangle)$ Bell
states. We obtain similar behavior for the Bell states as in earlier
results derived by using the Born-Markov approximation
\cite{FicTan06} on the same model. However, in the Class $|A
\rangle$ states the behavior is qualitatively different: under the
non-Markovian evolution we do not see sudden death of quantum
entanglement and subsequent revivals, except when the qubits are
sufficiently far apart.  (The existence of sudden death was first
reported for two qubits in two disjoint cavity electromagnetic
fields \cite{YuEbePRL}, and the dark period and revival were found
from calculations using the Born-Markov approximation
\cite{FicTan06}). For an initial Bell state, our findings based on
non-Markovian dynamics agree with those obtained under the
Born-Markov approximation. We provide explanations for such
differences of behavior both between these two classes of states and
between the predictions from the Markov and non-Markovian dynamics.
We also study the decoherence of this two-qubit system and find that
the decoherence rate in the case of one qubit initially in an
excited state does not change significantly with the qubits
separation whereas it does for the case when one qubit is initially
in the ground state. Furthermore, when the two qubits are close
together, the coherence of the whole system is preserved longer than
it does in the single qubit case or when the two qubits are far
apart.
\end{abstract}

\maketitle

\section{Introduction}

Investigation of quantum entanglement is both of practical and
theoretical significance: It is viewed as a basic resource for
quantum information processing (QIP) \cite{NielsenChuang} and it is
a basic issue in understanding the nature of nonlocality in quantum
mechanics \cite{EPR,Bell,GHZ}. However, even its very definition and
accurate characterization are by no means easy, especially for
multi-partite states (see, e.g.,
\cite{PeresBook,KarolBook,AlickiBook,Wootters,Bennett,Wer,VidWer}.)
Nonetheless there are useful criteria proposed for the separability
of a bipartite state, pure and mixed (e.g.,
\cite{Peres,Horod,Simon,Duan,Barnum,Brennen,HHH}), theorems proven
(e.g, \cite{Ruskai,AlickiHorod}), and new mathematical tools
introduced (e.g., \cite{Brody,Levay}), which add to advances in the
last decade of this new endeavor \cite{Knill}.

Realistic quantum systems designed for QIP cannot avoid interactions
with their environments, which can alter their quantum coherence and
entanglement properties. Thus quantum decoherence and
disentanglement are two essential obstacles to overcome for the
design of quantum computers and the implementation of QIP.
Environment-induced decoherence in the context of QIP has been under
study for over a decade \cite{PazZurRev} and studies of
environment-induced disentanglement has seen a rapid increase in
recent years. There are now  experimental proposals to measure
finite time disentanglement induced by a reservoir \cite{Franca}.
The relation between decoherence and disentanglement is an
interesting one because both are attributable to the decay of
quantum interference in the system upon the interaction with an
environment. (See, e.g., \cite{VKG,RajRendell,Diosi,Dodd,DoddHal})

In addition to the mathematical investigations mentioned above which
could provide rather general characterizations of quantum
entanglement, detailed studies of physical models targeting actual
designs of quantum computer components can add precious insight into
its behavior in concrete settings.  Two classes of models relevant
to condensed matter and atomic-optical QIP schemes are of particular
interest to us. The first class consists of  the quantum Brownian
motion model (QBM) and the spin-boson model (SBM). Quantum
decoherence has been studied in detail in both models, and results
on quantum disentanglement are also appearing (See
\cite{VKG,Diosi,Dodd,DoddHal} for QBM under high temperature,
negligible dissipation, and \cite{CHY} for an attempt towards the
full non-Markovian regimes.) The second class of models describes
atoms or ions in a cavity with a quantum electromagnetic field at
zero or finite temperature. The model consists of two two level
atoms (2LA) in an electromagnetic field (EMF). For a primary source
on this topic, read, e.g.. \cite{Agarwal}.  For a more recent
description of its dynamics under the Born-Markov approximation, see
the review of \cite{FicTan}.

\subsection{Two-atom entanglement via an electromagnetic field}

Quantum decoherence and entanglement between one 2LA and an EMF has
been treated by us and many other authors earlier \cite{AH,SADH} and
by Cummings and Hu recently towards the strong coupling regime
\cite{CumHu06}, which provide insight in how the atom-field
interaction affects their entanglement. There is recent report of
exact solutions found for a 2LA in an EMF using the underlying
pseudo-supersymmetry of the system \cite{SSG}. In the 2 atom-EMF
model, the two atoms can be assumed to interact only with its own
cavity EMF, or with a common EMF, and they can also interact with
each other. The noninteracting case of separate fields was first
studied by Yu and Eberly \cite{YuEbePRL,YuEbePRB} where `sudden
death' of quantum entanglement was sighted. The noninteracting case
of a common field was studied recently by Ficek and Tanas
\cite{FicTan06}, An {\it et al} \cite{AnWangLuo}. Quantum
decoherence of N-qubits in an electromagnetic field was studied by
Palma, Suominen and Ekert \cite{PSK}. For entanglement of ions in
cavities, see, e.g., \cite{BPH}.

For the purpose of quantum information processing, we have
emphasized earlier in our studies of quantum decoherence that it is
absolutely essential to keep track fully of the mutual influence of,
or the interplay between, the system and the environment. If one
chooses to focus only on the system dynamics, one needs to take into
consideration the back-action of the environment, and vice versa.

In our prior work \cite{AH,SADH}, we used the influence functional
formalism with a Grassmannian algebra for the qubits (system) and a
coherent state path integral representation for the EMF
(environment). Here, we employ a more standard operator method
through perturbation theory, because the assumption of an initial
vacuum state for the EMF allows a full resummation of the
perturbative series, thus leading to an exact and closed expression
for the evolution of the reduced density matrix of the two qubits.
This approach incorporates the back-action of the environment on the
system self-consistently and in so doing generates non-Markovian
dynamics. We shall see that these features make a fundamental
difference in the depiction of evolution of quantum entanglement in
the qubit dynamics.

\subsection{The importance of including back-action self-consistently}

Since quantum entanglement is a more delicate quantity to track down
than decoherence, an accurate description is even more crucial. For
this, one needs to pay extra attention to back-actions. For example,
in the case of two 2LA (system) in a cavity EMF (environment), the
two parties are equally important. This means that we should include
both the back-action of the field on the atoms, and the atoms on the
field. In a more complete treatment as attempted here, we obtain
results qualitatively different from earlier treatments where the
back-action is not fully included or properly treated \cite{FicTan}.
Some special effects like `sudden death' \cite{YuEbePRL} can in this
broader context be seen as consequences only of rather special
arrangements: Each atom interacting with its own EMF precludes the
fields from talking to each other and in turn cuts off the atoms'
natural inclination (by the dictum of quantum mechanics) to be
entangled. In effect, this is only a limiting case of the full
dynamics we explored here for the two-qubit entanglement via the
EMF. This limit corresponds to  the qubits being separated by
distances much larger than the correlation length characterizing the
total system. For a wide range of spatial separations within the
correlation length, entanglement is robust: Our results for the full
atom-field dynamics reveal that there is no sudden death.

\subsection{Non-Markovian dynamics from back-action}

It is common knowledge in nonequilibrium statistical mechanics
\cite{Zwanzig} that for two interacting subsystems the two ordinary
differential equations governing each subsystem can be written as an
integro-differential equation governing one such subsystem, thus
rendering its dynamics non-Markovian, with  the memory of the other
subsystem's dynamics registered in the nonlocal kernels (which are
responsible for the appearance of dissipation and noise should the
other subsystem possess a much greater number of degrees of freedom
and are coarse-grained in some way). Thus inclusion of back-action
self-consistently in general engenders non-Markovian dynamics.
Invoking the  Markov approximation as is usually done may lead to
the loss of valuable information, especially for quantum systems.
These assumptions need to be scrutinized carefully with due
consideration of the different time scales in the system and the
specific physical quantities that are of interest in the study.

For monitoring the evolution of quantum entanglement which is
usually a more delicate process than coherence, if one lacks
detailed knowledge of how the different important processes
interplay, our experience is that it is safer not to put in any ad
hoc assumption at the beginning (e.g., Markovian approximation, high
temperature, white noise) but to start with a first principles
treatment of the dynamics (which is likely non-Markovian) involving
all subsystems concerned and respecting full self-consistency. This
is because entanglement can be artificially and unknowingly
curtailed or removed in these ad hoc assumptions. What is described
here is not a procedural, but a substantive issue,  if one seeks to
coherently follow or manipulate any quantum system, as in QIP,
because doing it otherwise can generate quantitatively inaccurate or
even qualitatively wrong results.

Thus the inclusion of backreaction (which depends on the type of
coupling and the features of the environment) usually leads to
nonMarkovian dynamics \footnote{This does not preclude the
possibility that certain types of backreaction effects when included
leads to the same type of open system dynamics as the original
(test- field) dynamics, which, if already in the Lindblad form,
remains Markovian (e.g., renormalization of the coupling constants,
invoking a mean field approximation).}. Also, under extreme
conditions such as imposing infinite cutoff frequency and at high
temperatures, the dynamics of, say, a quantum harmonic oscillator
bilinearly coupled to an Ohmic bath can become Markovian
\cite{CalLeg83}.  Other factors leading to or effecting nonMarkovian
behavior include the choice of special initial conditions. For
example,  the factorizable initial condition introduces a fiducial
or special choice of time into the dynamics which destroys
time-homogeneity.

A word about terminology might be helpful here:   One usually refers
to Markovian dynamics as that governed by a master equation with
constant-in-time coefficients, i.e., described by a Linblad
operator,  and non-Markovian for all other types of dynamics. A more
restricted condition limits the definition of non-Markovian dynamics
to cases with non-trivial (nonlocal in time) integral kernels
appearing in the master equation. This more stringent definition
would refer to dynamics (depicted by master equations containing
coefficients which are) both time-homogeneous and
non-time-homogeneous as Markovian. We use the first and more common
convention of terminology, in which the master equation
(\ref{Lindbladlike}) which is local in time but non-time-homogeneous
would be nonMarkovian. The Markovian regime emerges in the limit
when the two qubits are far separated. This feature is similar to
the HPZ master equation \cite{HPZ} for quantum Brownian motion,
where Markovian (time-homogeneous) dynamics appears only in specific
limiting conditions (high temperature and ohmic distribution of
environmental modes, as alluded to above).

Our present study of the two qubit (2qb)- EMF system is also aimed
at addressing a common folklore, namely, that in quantum optics one
does not need to worry about non-Markovian effects. We will see that
there is memory effect residing in the off diagonal components of
the reduced density matrix for the 2 qubit system which comes from
virtual photon exchange processes mediated by the field and which
depends on the qubit separation. Perhaps the simplest yet strongest
reason for the need to take nonMarkovian effects seriously is that
results from the Markovian approximation are incomplete and lead to
qualitatively wrong predictions.

\subsection{Relation to prior work and organization of this paper}

In this paper we study the non-Markovian dynamics of  a pair of
qubits (2LA) separated in space by distance $r$ interacting with one
common electromagnetic field (EMF) through a Jaynes-Cummings type
interaction Hamiltonian. We use the concurrence function  as a
measure of quantum entanglement between the two qubits.  The same
model was studied before in detail by Ficek and Tanas \cite{FicTan}
using the Born-Markov approximation.
In a more recent paper \cite{FicTan06} they show the existence of
dark periods and revival of quantum entanglement in contrast to the
findings of Yu and Eberly which assumes two qubits in disjoint EMFs.

Our calculation makes a weak coupling assumption and ignores the
two- and higher- photon- exchanges between the qubits, but it makes
no Born or Markov approximation. We derive a non-Markovian master
equation, which differs from the usual one of the Lindblad type: it
contains extra terms that correspond to off-diagonal elements of the
density matrix propagator. We concentrate on two classes of states,
superpositions of highest and lowest energy states and the usual
antisymmetric $|- \rangle$ and symmetric $|+\rangle $  Bell states
\cite{Bell} and observe very different behavior. These are described
in detail in the Discussions section. The difference between our
results and that of Ref. \cite{FicTan06} highlights the effect of
non-Markovian (with memory) evolution of quantum entanglement. In
short, we find similar behavior in the Class B (Bell) states  but
qualitative different behavior in the evolution of Class A states.
Ref \cite{FicTan06} found that their evolution leads generically to
sudden death of entanglement and a subsequent revival. In our more
complete treatment of the atom-field dynamics we indeed see the
former effect present for large values of the inter-qubit distances.
However, sudden death is absent for short distances, while there is
no regime in which a revival of entanglement can take place. This
calls for caution.

Another set of papers close to our work reported here is that of
\cite{JakJam} who considered two 2LA in an infinite temperature
field bath. When the atoms are separated at large distance the
authors assume that they are located inside two independent baths. (
The severance of the field is subject to the same criticism above: A
small but finite quantum entanglement cannot be equated to zero
because the small amount can later grow.) For these conditions and
under the Markovian approximation, the time evolution of the
two-atom system is given by the ergodic dynamical semi-group. They
ignore without justification the effect of distance on the
interaction between the qubits.  A paper of interest not directly
related to the present model but which does show the dependence of
the disentanglement rate on distance, like ours reported here, is
that by Roszak and Machinikowski \cite{RosMac}. They consider a
system of excitons with different coupling, and with a very
different infrared behavior of the bath modes. The latter seems not
to be relevant to the two atoms' case here.

This paper is organized as follows: Section 2 contains the main
derivation. We write down the Hamiltonian for two 2-level atoms
(2LA) interacting with a common electromagnetic field (EMF) at zero
temperature, and we compute the relevant matrix elements for the
propagator of the total system by resummation of the perturbative
series (Appendix A). We then determine the evolution of the reduced
density matrix of the atoms, which is expressed in terms of seven
functions of time. We compute these functions using an approximation
that amounts to keeping the contribution of the lowest loop order
for the exchange of photons between the qubits. In Section 3 we
examine the evolution of the reduced density matrix for two classes
of initial states. We then describe the time evolution of quantum
entanglement with spatial separation dependence in these states via
the concurrence plotted for some representative cases. In Section 4,
we study the decoherence of this system when the two qubits are
initially disentangled. We consider two cases that correspond to one
of the qubits being initially in a vacuum state and in an excited
state. We compare these results with the single qubit cases and
highlight the lessening of decoherence due to the presence of other
qubit(s). In Section 5 we discuss and compare our results on
disentanglement with the work of Yu and Eberly for two 2LA in
separate EMF baths, and with the work of Fizek and Tanas on two 2LA
in a common EMF bath under the Born-Markov approximation. We
identify the point of departure of quantum dynamics under the
Markovian approximation from the full non-Markovian dynamics and
thereby demonstrate the limitations of the Born-Markov
approximation. Finally, we discuss the domain of validity of the
rotating wave approximation in describing these systems. In Appendix
C we sketch an alternative derivation through the Feynman-Vernon
influence functional technique, in which Grassmann variables are
employed for the study of the atomic degrees of freedom.

\section{Two-Atoms interacting via a common Electromagnetic Field}

\subsection{The Hamiltonian}

We consider two 2-level atoms (2LA),  acting as two qubits, labeled
1 and 2, and an electromagnetic field described by the free
Hamiltonian $H_0$
\begin{eqnarray}
\hat{H}_0 = \hbar \sum_{\bf k} \omega_{\bf k}\hat{b}^{\dagger}_{\bf
k} \hat{b}_{\bf k} +\hbar\omega_o  \hat{S}_+^{(1)} \hat{S}_-^{(1)}
+\hbar\omega_o \hat{S}_+^{(2)} \hat{S}_-^{(2)}
\end{eqnarray}
where $\omega_\bk$ is the frequency of the $\bk^{\mbox{th}}$
electromagnetic field mode and $\omega_o$ the atomic frequency
between the two levels of the atom, assumed to be the same for the
two atoms. The electromagnetic field creation~(annihilation)
operator is $b_\bk^+$~($b_\bk$), while $S_+^{(n)}$~($S_-^{(n)}$) are
the spin raising~(lowering) operators for the $n^{\mbox{th}}$ atom.
We will define the pointing vector from $1$ to $2$ as $\br =\br_2
-\br_1$ and we will assume without loss of generality that $\br_1 +
\br_2 = 0$.

 The two 2LAs do not interact with each other directly
but only through the common electromagnetic field via the
interaction Hamiltonian
\begin{eqnarray}
\hat{H}_I = \hbar\sum_{\bf k} g_{\bf k}\left( \hat{b}^{\dagger}_{\bf
k} ( e^{-i\bk\cdot\br/2} \hat{S}_{-}^{(1)} + e^{i\bk\cdot\br/2}
\hat{S}_{-}^{(2)} ) + b_{\bf k} ( e^{i\bk\cdot\br/2}
\hat{S}_{+}^{(1)} + e^{-i\bk\cdot\br/2} \hat{S}_{+}^{(2)} ) \right),
\label{Hint}
\end{eqnarray}
where $g_{\bf k } = \frac{\lambda}{\sqrt{\omega_{\bf k}}}$,
$\lambda$ being the coupling constant. We have assumed that the
dipole moments of the atoms are parallel. The total Hamiltonian of
the atom-field system is
\begin{equation}
\hat{H} = \hat{H}_0 + \hat{H}_I.
\end{equation}

\subsection{Perturbative expansion and resummation}

We assume that at $t = 0$ the state of the combined system of
atoms+field is factorized and that the initial state of the EMF is
the vacuum $|O \rangle$. For this reason we need to identify the
action of the evolution operator $e^{-i\hat{H}t}$ on vectors of the
form $|O \rangle \otimes | \psi \rangle$, where $|\psi \rangle$ is a
vector on the Hilbert space of the two 2LA's.

For this purpose, we use the resolvent expansion of the Hamiltonian
\begin{eqnarray}
e^{-i\hat{H}t} = \int \frac{dE e^{-iE t}}{E - \hat{H} + i \eta}
\end{eqnarray}
and we expand
 \begin{eqnarray}
 (E - \hat{H})^{-1} = (E - \hat{H}_0)^{-1} + (E - \hat{H}_0)^{-1} \hat{H}_I (E - \hat{H}_0)^{-1}
\nonumber \\ + (E - \hat{H}_0)^{-1} \hat{H}_I (E - \hat{H}_0)^{-1}
\hat{H}_I (E - \hat{H}_0)^{-1} + \ldots \label{expansion}
\end{eqnarray}

Of relevance for the computation of the reduced density matrix of
the two qubits are matrix elements of the form $\langle z; i',j'|
 (E - \hat{H})^{-1}| O; i,j \rangle$, where $z$ refers to a coherent state of the EM
 field and $i, j = 0,1$, the value $i = 0$ corresponding to the
 ground state of a single qubit and $i = 1$ to the excited state.
  We compute the matrix elements above through the
perturbation expansion (\ref{expansion}).
 It turns out that we can
effect a resummation of the perturbative series and thus obtain an
exact expression for the matrix elements--see Appendix A for details
of the resummation.

The non-vanishing matrix elements are the following
\begin{eqnarray}
\langle z; 0,0| (E - \hat{H})^{-1}| O; 0, 1 \rangle &=& \sum_{\bf k}
\frac{g_{\bf k} z^*_{\bf k} e^{ i \frac{{\bf k} \cdot {\bf
r}}{2}}}{(E - \w_o - \alpha(E) - \beta(E,r)e^{i {\bf k} \cdot {\bf
r}})(E - \omega_{\bf k})}
\\
\langle z; 0,1| (E - \hat{H})^{-1}| O; 0, 1 \rangle &=&
\frac{1}{2}\left[
\frac{1}{E - \w_o - \alpha(E) - \beta(E,r)} \right. \nonumber \\
&+& \left. \frac{1}{E - \Omega - \alpha(E) + \beta(E,r)} \right]
 \\
\langle z; 1,0| (E - \hat{H})^{-1}| O; 0, 1 \rangle &=&
\frac{1}{2}\left[
\frac{1}{E - \w_o - \alpha(E) - \beta(E,r)} \right. \nonumber \\
 &-& \left. \frac{1}{E - \w_o - \alpha(E) + \beta(E,r)} \right]
 \\
\langle z; 0,0| (E - \hat{H})^{-1}| O; 0, 0 \rangle &=& E^{-1}
\\
\langle z; 1,1| (E - \hat{H})^{-1}| O; 1, 1 \rangle &=& \frac{1}{E -
2\w_o - 2 \alpha (E - \w_o)- f(E,r)} \\
\langle z; 0,0| (E - \hat{H})^{-1}| O; 1, 1 \rangle &=& \sum_{{\bf k
k'}} \frac{\hat{H}_{{\bf k k'}} z^*_{{\bf k}} z^*_{{\bf k}}}{E -
2\w_o - 2 \alpha (E - \w_o)- f(E,r)}\\
 \left( \begin{array}{cc} \langle z; 0,1| (E - \hat{H})^{-1}| O; 1, 1 \rangle \\
\langle z; 1, 0| (E - \hat{H})^{-1}| O; 1, 1 \rangle \end{array}
\right) &=& \sum_{\bf k k'} \frac{g_{\bf k'} z^*_{\bf k}}{(E - 2
\w_o)(E - \w_o - \omega_{\bf k'})} \left( \begin{array}{cc}
e^{-i\frac{{\bf k} \cdot {\bf r}}{2}} \\ e^{i\frac{{\bf k} \cdot
{\bf r}}{2}}
\end{array} \right) (1
-L)^{-1}_{\bf kk'}
\end{eqnarray}

In the equations above the functions $\alpha(E), \beta(E,r)$ are
\begin{eqnarray}
\alpha(E) :&=& \sum_{\bf k} \frac{g_{\bf k}^2}{E - \omega_{\bf k}}
\\
\beta(E,r) :&=& \sum_{\bf k} \frac{g_{\bf k}^2}{E - \omega_{\bf k}}
e^{i{\bf k} \cdot {\bf r} }.
\end{eqnarray}

The definitions of the kernel $H_{\bf k k'}$ and of the function $f$
involve complicated expressions. However, the term involving $H_{\bf
kk'}$ does not contribute to the evolution of the reduced density
matrix, while the function $f(E,r)$ is of order $\lambda^4$ and it
can be ignored in the approximation we effect in Sec. II.E. Thus for
the purpose of this investigation, the explicit definitions of $H$
and $f$ are not needed, and hence not given here.

Finally, the matrix $L$ is defined as
\begin{eqnarray}
L := \left( \begin{array}{cc} \Xi & \Theta \\ \bar{\Theta} &
\bar{\Xi}
\end{array} \right),
\end{eqnarray}
where $\Xi$ and $\Theta$ are matrices on the space of momenta
\begin{eqnarray}
\Xi_{\bf k k'} = \frac{1}{E - \Omega - \omega_{\bf k}} \left(
\alpha(E - \omega_{\bf k}) \delta_{\bf kk'} + g_{\bf k} g_{\bf k'}
(\frac{1}{E-2 \Omega} + \frac{e^{i ({\bf k} - {\bf k'}) \cdot {\bf
r}}}{E - \omega_{\bf k} - \omega_{\bf k'}} )\right) \label{Xi}
\\
\Theta_{\bf k k'}  = \frac{1}{E - \Omega - \omega_{\bf k}} \left(
\beta(E-\omega_{\bf k}, r) \delta_{\bf kk'} + g_{\bf k} g_{\bf k'}
(\frac{1}{E-2 \Omega} + \frac{ 1}{E - \omega_{\bf k} - \omega_{\bf
k'}} )\right), \label{Theta}
\end{eqnarray}
and the overbar denotes complex conjugation.

\subsection{The matrix elements of the propagator}

The next step is to Fourier transform the  matrix elements of the
resolvent  in order to obtain the matrix elements of the evolution
operator. Explicitly,
\begin{eqnarray}
\langle z; 0, 0| e^{-i\hat{H}t}| O; 0, 1 \rangle &=& \sum_{\bf k}
e^{i
{\bf k}\cdot {\bf r}/2} z^*_{\bf k} s_{\bf k}(t) \label{00}\\
\langle z; 0, 1| e^{-i\hat{H}t}| O; 0, 1 \rangle &=& \int \frac{dE
e^{-iEt}}{2} \left[ \frac{1}{E - \w_o - \alpha(E) - \beta(E,r)}
\right. \nonumber \\ &+& \left.\frac{1}{E - \w_o - \alpha(E) +
\beta(E,r)} \right]
= : v_+(t) \label{iv+}\\
\langle z; 1, 1| e^{-i\hat{H}t}| O; 0, 1 \rangle &=& \int \frac{dE
e^{-iEt}}{2} \left[ \frac{1}{E - \w_o - \alpha(E) - \beta(E,r)}
\right. \nonumber \\ &-&
\left. \frac{1}{E - \w_o - \alpha(E) + \beta(E,r)} \right] =: v_-(t) \label{iv-}\\
 \langle z; 0, 0| e^{-i\hat{H}t}| O; 0, 0 \rangle &=& 1 \\
 \langle z; 1, 1| e^{-i\hat{H}t}| O; 1, 1 \rangle &=& \int \frac{dE
 e^{-iEt}}{E - 2 \w_o - 2 \alpha(E - \w_o) -f(E, r)} =: u(t) \label{iu}\\
 \langle z; 0, 0| e^{-i\hat{H}t}| O; 1, 1 \rangle &=& \int dE e^{-iEt}  \sum_{\bf k k'}
 \frac{ \hat{H}_{\bf k k'} z^*_{\bf k} z^*_{\bf k'} }{ E - 2 \w_o - 2 \alpha(E - \w_o)
 -f(E, r)} \\
\left( \begin{array}{cc}  \langle z; 0, 1| e^{-i\hat{H}t}| O; 1, 1
\rangle
\\ \langle z; 1, 0| e^{-i\hat{H}t}| O; 1, 1 \rangle \end{array} \right)
&=& \sum_{\bf k} z^*_{\bf k} \left( \begin{array}{cc}
e^{-i\frac{{\bf k} \cdot {\bf r}}{2}} \nu_{\bf k}(t) \\
e^{i\frac{{\bf k} \cdot {\bf r}}{2}} \nu'_{\bf k}(t) \label{11}
\end{array} \right),
\end{eqnarray}
where we defined the functions $s_{\bf k}(t), \nu_{\bf k}(t),
\nu_{\bf k}'(t)$ as
\begin{eqnarray}
s_{\bf k}(t)  &=&  \int \frac{dE e^{-iEt}}{(E-\w_o - \alpha(E) -
\beta(E,r) e^{i {\bf k} \cdot {\bf r}})(E -
\omega_{\bf k})} \label{sk}\\
\left( \begin{array}{cc} \nu_{\bf k}(t) \\ \nu'_{\bf k}(t)
\end{array} \right) &=& \int \frac{dE e^{-iEt}}{E - 2 \w_o} \sum_{\bf k'} (1 - L)_{\bf kk'} \left(
\begin{array}{cc} \frac{g_{\bf k'}}{E - \w_o - \omega_{\bf k'}} \\ \frac{g_{\bf k'}}{E - \w_o - \omega_{\bf
k'}} \end{array} \right) \label{nu}
\end{eqnarray}

\subsection{The reduced density matrix}

We next compute the elements of the reduced density matrix for the
qubit system by integrating out the EM field degrees of freedom
\begin{eqnarray}
\rho^{ij}_{i'j'}(t) = \sum_{i_0,j_0,i'_0,j'_0}
\rho^{i_0,j_0}_{i'_0,j'_0} (0) \int [dz] [dz^*] \langle O;
i'_0j'_0|e^{i\hat{H}t}| z; i', j' \rangle \langle
z,i,j|e^{-i\hat{H}t} |O, i_0,j_0 \rangle, \label{rdm}
\end{eqnarray}
where $[dz]$ is the standard Gaussian integration measure for the
coherent states of the EM field.

Substituting Eqs. (\ref{00}-\ref{11}) into (\ref{rdm}) we obtain
through a tedious but straightforward calculation the elements of
the reduced density matrix

\begin{eqnarray} \rho^{I}_{I}(t) &=& \rho^{I}_{I}(0) |u|^2(t)
\label{1111}
\\
\rho^{11}_{01}(t) &=& \rho^{11}_{01}(0) u(t) v^*_+(t) +
\rho^{11}_{10}(0) u(t) v^*_-(t)
\\
\rho^{11}_{10}(t) &=& \rho^{11}_{10}(0) u(t) v^*_+(t) +
\rho^{11}_{01}(0) u(t) v^*_-(t)
\\
\rho^{I}_{00}(t) &=& \rho^{11}_{00} u(t)
\\
 \rho^{01}_{00}(t) &=& \rho^{01}_{00}(0) v_+(t) + \rho^{10}_{00}(0)
v_-(t) + \rho^{11}_{01}(0) \mu_1(t) + \rho^{11}_{10}(0) \mu_2(t)
\\
\rho^{10}_{00}(t) &=& \rho^{10}_{00}(0) v_+(t) + \rho^{10}_{00}(0)
v_-(t) + \rho^{11}_{01}(0) \mu^*_2(t) + \rho^{11}_{10}(0) \mu^*_1(t)
\\
\rho^{01}_{01}(t) &=& \rho^{01}_{01}(0) |v_+|^2(t) +
\rho^{01}_{10}(0) v_+(t) v^*_-(t) +  \rho^{10}_{10}(0) |v_-|^2(t)
\nonumber \\ &+& \rho^{10}_{01}(0)  v_-(t) v^*_+(t) +
\rho^{11}_{11}(0) \kappa_1(t)
\\
\rho^{01}_{10}(t) &=& \rho^{01}_{10}(0) |v_+|^2(t) +
\rho^{10}_{01}(0) |v_-|^2(t) +  \rho^{01}_{01}(0)  v_+(t)
v_-^*(t)\nonumber \\ & +& \rho^{10}_{10}(0) v_-(t) v^*_+(t) +
\rho^{11}_{11}(0) \kappa_2(t)
\\
\rho^{00}_{00}(t) &=& 1 - \rho^{11}_{11}(t) - \rho^{01}_{ 01}(t) -
\rho^{10}_{10}(t) \label{0000}
\end{eqnarray}
where
\begin{eqnarray}
\mu_1(t) &=& \sum_{\bf k} g_{\bf k} \nu_{\bf k}(t) s^*_{\bf k}(t)
\label{defm1}
\\
\mu_2(t) &=& \sum_{\bf k} g_{\bf k} \nu_{\bf k}(t) s^*_{\bf k}(t)
e^{-i {\bf k} \cdot {\bf r}} \label{defm2}\\
\kappa_1(t) &=& \sum_{\bf k}  |\nu_{\bf k}|^2(t)  \label{defk1}\\
\kappa_2(t) &=& \sum_{\bf k} \nu_{\bf k}(t) \nu'^*_{\bf k}(t) e^{- i
{\bf k} \cdot {\bf r}}, \label{defk2}
\end{eqnarray}
and the functions $u(t), v_{\pm}(t)$ were defined in Eqs.
(\ref{iu}), (\ref{iv+}) and (\ref{iv-}).

\subsection{Explicit forms for the evolution functions}

Eqs. (\ref{1111}-\ref{0000}) provide an {\em exact} expression for
the evolution of the reduced density matrix for the system of two
qubits interacting with the EM field in the vacuum state. The
evolution is determined by seven functions of time $u, v_{\pm},
\kappa_{1,2}, \mu_{1,2}$, for which we have provided the full
definitions. To study the details of the qubits' evolution we must
obtain explicit forms for the functions above. For analytic
expressions, we recourse to an approximation: Assuming weak coupling
($\lambda^2 << 1$), we ignore the contribution of all processes that
involve the exchange of two or more photons between the two qubits.

\subsubsection{The functions $u, v_{\pm}$}

With the approximation above, the contribution of the function $f$
drops out from the definition of $u$. Thus we obtain
\begin{eqnarray}
 u(t) &=& \int \frac{dE
 e^{-iEt}}{E - 2 \w_o - 2 \alpha(E - \w_o) } \\
 v_{\pm} &=& \int dE e^{-iEt}
\frac{1}{2}\left[ \frac{1}{E - \w_o - \alpha(E) - \beta(E,r)} \pm
\frac{1}{E - \w_o - \alpha(E) + \beta(E,r)} \right].
\end{eqnarray}

We evaluate these expressions using an additional approximation. In
performing the Fourier transform, we only keep the contribution of
the poles in the integral and ignore that coming from a branch-cut
that appears due to the presence of a logarithm in the exact
expression of $\alpha(E)$--see Ref. \cite{AH} for details. We then
obtain
\begin{eqnarray}
u(t) &=& e^{ - 2 i \omega_o t - 2 \Gamma_0 t} \label{u}
\\
v_{\pm}(t) &=& \frac{e^{- i \omega_o t - \Gamma_0 t}}{2} \left( e^{
- i \sigma t - \Gamma_r t} \pm e^{i \sigma t + \Gamma_r t} \right).
\label{v+-}
\end{eqnarray}

In the equations above, we effected a renormalization of the
frequency $\w_o$ by a constant divergent term-- see \cite{AH}. The
parameters $\g_0, \Gamma_r$ and $\sigma(r) $ are defined as
\begin{eqnarray}
\g_0 := - Im \, \alpha(\w_o) \\
-\sigma(r) + i \Gamma_r := \beta(\w_o, r),
\end{eqnarray}
and they read explicitly
\begin{eqnarray}
\g_0 &=& \frac{\lambda^2 \w_o}{2 \pi} \\
\Gamma_r &=& \frac{\lambda^2 \sin \omega_o r }{2 \pi r} \\
\s(r) &=& \frac{\lambda^2}{2 \pi^2 r} \left[ - \cos
\omega_0 r [\frac{\pi}{2} - Si(\omega_0 r)] \right.  \nonumber \\
 &+& \left. \sin \omega_0r [\log(e^{\gamma} \omega_0 r) +
\int_0^{\omega_0r} dz \, \frac{1 - \cos z}{z}] \right],
\end{eqnarray}
where $Si$ is the sine-integral function.

The term $\s(r)$ is a frequency shift caused by the vacuum
fluctuations. It breaks the degeneracy of the two-qubit system and
generates an effective dipole coupling between the qubits. At the
limit $r \rightarrow 0$, this term becomes infinite. One should
recall however that the physical range of $r$ is always larger than
$a_B$, the Bohr radius of the atoms. As $r \rightarrow \infty$,
$\s(r) \rightarrow 0$.

The constant $\g_0$ corresponds to the rate of emission from
individual qubits. It coincides with the rate of emission obtained
from the consideration of a single qubit interacting with the
electromagnetic field. The function $\Gamma_r$ is specific to the
two-qubit system. It arises from  Feynman diagrams that involve an
exchange of photons between the qubits. Heuristically, it expresses
the number of virtual photons
 per unit time exchanged between the qubits. As such, $\Gamma_r^{-1}$ is
 the characteristic time-scale for the exchange of information
 between the qubits.  As $r \rightarrow 0$,  $\Gamma_r \rightarrow
 \g_0$ and as $r \rightarrow \infty$, $\Gamma_r \rightarrow 0$. Note
 that the ratio $\g/\g_0 =
 \frac{\sin \omega_o r}{\omega_o r}$, while smaller than unity, is of
 the order of unity as long as $r$ is not much larger than
 $\omega_o^{-1}$.

It is interesting to note that $\Gamma_r = 0$ for  $r = n \pi
\omega_0^{-1}$, where $n$ an integer. This is a resonant behaviour,
similar to that of a classical oscillating dipole when $r = n
\lambda/2$, where $\lambda$ is the oscillation wavelength.

\subsubsection{The functions $\kappa_{1,2}(t)$}
We first compute the functions $\nu_{\bf k}, \nu'_{\bf k}$ of Eq.
(\ref{nu}) keeping terms up to second loop order
\begin{eqnarray}
\left( \begin{array}{cc} \nu_{\bf k}(t) \\ \nu'_{\bf k}(t)
\end{array} \right) &=& \int \frac{dE e^{-iEt}}{E - 2 \omega_o} \sum_{\bf k'}  \frac{g_{\bf k'}}
{E - \omega_o - \omega_{\bf k'}}\left(
\begin{array}{cc} \delta_{\bf kk'} + \Xi_{\bf kk'} + \Theta_{\bf kk'} \\\delta_{\bf kk'} +
 \bar{\Xi}_{\bf kk'} + \bar{\Theta}_{\bf kk'}\\  \end{array}
\right),
\end{eqnarray}
where $\Xi$ and $\Theta$ are given by Eqs. (\ref{Xi}) and
(\ref{Theta}).

The summation over ${\bf k'}$ yields within an order of $\lambda^5$
\begin{eqnarray}
\left( \begin{array}{cc} \nu_{\bf k}(t) \\ \nu'_{\bf k}(t)
\end{array} \right) &=& \int \frac{dE e^{-iEt}}{E - 2 \omega_o} \frac{g_{\bf k}}{E - \omega_o - \omega_{\bf k}} \left[ 1 +
\frac{ \alpha(E - \omega_{\bf k}) + \beta(E - \omega_{\bf k})}{E -
\omega_o - \omega_{\bf k}}\right] \left[ 1  + 2 \frac{\alpha(E -
\omega_o)}{E - 2 \omega_o} \right]
 \nonumber \\ &\times& \left( \begin{array}{cc} 1 + \sum_{\bf
k'} \frac{g_{\bf k'}^2 (1 + e^{i ({\bf k} - {\bf k'}) \cdot {\bf
r}})} {(E -
\omega_{\bf k} - \omega_{\bf k'}) (E - \omega_o - \omega_{\bf k'})} \\
1 + \sum_{\bf k'} \frac{g_{\bf k'}^2 (1 + e^{-i ({\bf k} - {\bf k'})
\cdot {\bf r}})} {(E - \omega_{\bf k} - \omega_{\bf k'}) (E -
\omega_o - \omega_{\bf k'})} \end{array} \right). \label{nuF}
\end{eqnarray}

The terms in brackets in the first line of the equation above can be
absorbed in the leading-order denominators--see Eq. (\ref{nnn}). The
term in the second line, however, only gives rise to a
(time-independent) multiplicative term  of the form $1 +
O(\lambda^2)$. Hence, if we  keep the leading order terms in the
expression of $\nu_{\bf k}$, we may  ignore this term. Then
$\nu_{\bf k} = \nu'_{\bf k}$, and within an error  of order
$\lambda^5$
\begin{eqnarray}
\nu_{\bf k}(t) = g_{\bf k} \int \frac{dE e^{-iEt}}{[E - 2 \omega_o -
2 \alpha (E - \omega_o)][E - \omega_o - \omega_{\bf k} - \alpha(E -
\omega_{\bf k}) - \beta(E - \omega_{\bf k}, r))]}. \label{nnn}
\end{eqnarray}

Using the same approximation as in Sec. II.E.1 for the Fourier
transform, we obtain
\begin{eqnarray}
\nu_{\bf k}(t) = g_{\bf k} \frac{e^{ - i \Omega t - \Gamma_0
t}}{\omega - \omega_{\bf k} - \sigma -i \Gamma_0 + i \Gamma_r}
\left( e^{- i \omega_0 t - \Gamma_0t} - e^{ -i \omega_{\bf k} t - i
\sigma t - \Gamma_r t} \right)
\end{eqnarray}

We then substitute the expression above for $\nu_{\bf k}$ into Eqs.
(\ref{defk1}) and (\ref{defk2}) to get
\begin{eqnarray}
\kappa_1(t) = \frac{\lambda^2}{2 \pi^2} e^{-2 \Gamma_0t}
\int_0^{\infty} k dk \, \frac{e^{-2 \Gamma_0 t} + e^{ - 2\Gamma_r t}
- 2 e^{ - (\Gamma_0 + \Gamma_r) t} \cos [ (\omega_o - k -
\sigma)t]}{(k - \omega_o +
\sigma)^2 + (\Gamma_0 - \Gamma_r)^2} \\
\kappa_2(t) = \frac{\lambda^2}{2 \pi^2 r} e^{-2 \Gamma_0t}
\int_0^{\infty} dk \,  \frac{e^{-2 \Gamma_0 t} + e^{ - 2\Gamma_r t}
- 2 e^{ - (\Gamma_0 + \Gamma_r) t} \cos [ (\omega_o - k -
\sigma)t]}{(k - \omega_o + \sigma)^2 + (\Gamma_0 - \Gamma_r)^2} \sin
kr
\end{eqnarray}

For $\w_o t >> 1$, it is a reasonable approximation to substitute
the Lorentzian with a delta function. Hence,
\begin{eqnarray}
\kappa_1(t) = \Gamma_0 \kappa(t) \label{kappa1} \\
\kappa_2(t) = \Gamma_r \kappa(t) \label{kappa2}
\end{eqnarray}
where
\begin{eqnarray}
\kappa(t) \simeq \frac{1}{\Gamma_0 - \Gamma_r} e^{-2 \Gamma_0t}
(e^{-\Gamma_0 t} - e^{- \Gamma_rt})^2  \label{kappa}.
\end{eqnarray}

\subsubsection{The functions $\mu_{1,2}$}

We first compute the functions $s_{\bf k}(t)$ of Eq. (\ref{sk})
\begin{eqnarray}
s_{\bf k}(t) = \frac{ e^{-i \w_o t -  \Gamma_0 t -(\Gamma_r + i
\sigma) e^{i {\bf k} \cdot{\bf r}} t} - e^{- i \omega_{\bf
k}t}}{\w_o - \omega_{\bf k} - i \Gamma_0 + (\sigma - i \Gamma_r)
e^{i {\bf k} \cdot{\bf r}}}
\end{eqnarray}
Substituting into Eqs. (\ref{defm1}) and (\ref{defm2}) we obtain
\begin{eqnarray}
\mu_1(t) = \frac{\lambda^2}{4 \pi^2} \int_{-1}^{1} d \xi \; \int k
dk \; e^{ - i \w_o t - \Gamma_0 t}\frac{e^{- i \omega_0 t -
\Gamma_0t} - e^{ -i \omega_{\bf k} t - i \sigma t - \Gamma_r
t}}{\Omega - \omega_{\bf k} - \sigma -i \Gamma_0 + i \Gamma_r}
\nonumber\\
\times \frac{ e^{i \w_o t -  \Gamma_0 t -(\Gamma_r - i \sigma) e^{-i
{\bf k} \cdot{\bf r}} t} - e^{ i \omega_{\bf k}t}}{\w_o -
\omega_{\bf k} + i \Gamma_0 + (\sigma + i \Gamma_r) e^{-i {\bf k}
\cdot{\bf r}}}\\
\mu_2(t) = \frac{\lambda^2}{4 \pi^2} \int_{-1}^{1} d \xi \; \int k
dk e^{-i {\bf k} \cdot {\bf r}} \; e^{ - i \w_o t - \Gamma_0
t}\frac{e^{- i \omega_0 t - \Gamma_0t} - e^{ -i \omega_{\bf k} t - i
\sigma t - \Gamma_r t}}{\w_o - \omega_{\bf k} - \sigma -i \Gamma_0 +
i \Gamma_r}
\nonumber\\
\times \frac{ e^{i \w_o t -  \Gamma_0 t -(\Gamma_r - i \sigma) e^{-i
{\bf k} \cdot{\bf r}} t} - e^{ i \omega_{\bf k}t}}{\w_o -
\omega_{\bf k} + i \Gamma_0 + (\sigma + i \Gamma_r) e^{-i {\bf k}
\cdot{\bf r}}}.
\end{eqnarray}
An approximate evaluation of the $\xi$-integral followed by the
further approximation $\frac{1}{x+ i \epsilon} \simeq i \pi
\delta(x)$, gives an estimation of the leading order contribution
\begin{eqnarray}
\mu_1(t) = \g_0 [\mu(t) + i \nu(t)] \label{mu1}\\
\mu_2(t) = \g_r [\mu(t) + i \nu(t)] \label{mu2}
\end{eqnarray}
where $\mu + i \nu$ is the complex-valued function
\begin{eqnarray}
\mu(t) + i \nu(t)=   \simeq \frac{1 }{\Gamma_0 + \frac{ 2 \sin \w_o
r}{\w_o r} \Gamma_r - i \sigma ( 1 + 2 \frac{\sin \w_o r}{\w_o r})}
e^{- i \w_o t - \Gamma_0 t}\nonumber
\\ \times ( e^{ - \Gamma_0 t} - e^{- \Gamma_rt} )( e^{ - \Gamma_0 t
- 2 \frac{\sin \w_o r}{\w_o r} [\Gamma_r - i \sigma] t} - e^{ i
\sigma t} ). \label{munu}
\end{eqnarray}

\subsection{The master equation}

Given the explicit form of the functions computed in Sec. II.E, we
write the evolution equations in the following form
\begin{eqnarray}
\rho^{I}_{I} (t) &=& e^{-4 \Gamma_0 t} \rho^{I}_{I}(0) \label{dmp1}\\
\rho^{I}_{O}(t) &=& e^{ - 2 i \omega_o t - 2 \Gamma_0 t}
\rho^{I}_{O}(0)
\\
 \rho^{I}_-(t) &=& e^{-i \omega_o t - 3 \Gamma_0 t + i
\sigma t +
\Gamma_r t} \rho^{I}_-(0) \\
\rho^{I}_+ (t) &=& e^{-i \omega_o t - 3 \Gamma_0 t - i \sigma t
-\Gamma_r t} \rho^{I}_+(0) \\
\rho^-_O(t) &=& e^{-i \omega_ot -\Gamma_0t + i \sigma  + \Gamma_r t}
\rho^-_O(0) + i (\Gamma_0 + \Gamma_r)  \nu(t) \rho^{I}_+(0) +
(\Gamma_0 - \Gamma_r)  \mu(t) \rho^{I}_-(0) \\
\rho^+_{O}(t) &=& e^{-i \omega_ot -\Gamma_0t - i \sigma  - \Gamma_r
t} \rho^+_{O}(0) + (\Gamma_0 + \Gamma_r)  \mu(t) \rho^{I}_+(0) +
i (\Gamma_0 - \Gamma_r) \nu(t) \rho^{I}_-(0) \\
\rho^+_+(t) &=& e^{-2 \Gamma_0 t - 2 \Gamma_r t} \rho^+_+(0) +
(\Gamma_0 + \Gamma_r) \kappa(t) \rho^{I}_{I}(0) \\
\rho^-_-(t) &=& e^{-2 \Gamma_0 t - 2 \Gamma_r t} \rho^-_-(0) +
(\Gamma_0 - \Gamma_r) \kappa(t) \rho^{I}_{I}(0) \label{rss}\\
\rho^+_-(t) &=& e^{2 i \sigma t - 2 \Gamma_0 t} \rho^+_-(0).
\label{dmpf}
\end{eqnarray}

In the equations above we wrote the density matrix in a basis
defined by $|I \rangle, |O\rangle, |+\rangle, |-\rangle$, where
\begin{eqnarray} |+ \rangle = \frac{1}{\sqrt{2}} \left( |01 \rangle +
|10 \rangle
\right) \\
|- \rangle = \frac{1}{\sqrt{2}} \left( |01 \rangle - |10 \rangle
\right).
\end{eqnarray}

We note that in this approximation, the diagonal elements of the
density matrix propagator (i.e. the ones that map $\rho^a_b(0)$ to
$\rho^a_b(t)$ where $a,b \in \{O, +, -, I\}$) decay exponentially
(except for the $\rho^{O}_{O}$ which is functionally dependent on
the other matrix elements due to normalization). We shall see in
Sec. V that this behavior is in accordance with the Born-Markov
approximation. The situation is different for the off-diagonal
elements of the density matrix propagator, i.e. the ones that map
$\rho^a_b(0)$ to $\rho^{a'}_{b'}(t)$ for $a \neq a'$ and $b \neq
b'$. They are given by more complex functions of time and they
differ from the ones predicted by the Born-Markov approximation.

As we have the solution $\rho(t) = M_t[\rho(0)]$, where $M_t$ is the
density matrix propagator defined by Eqs. (\ref{dmp1}-\ref{dmpf}),
we can identify the master equation through the relation
$\dot{\rho}(t) = \dot{M}_t[ M^{-1}_t[\rho(t)]]$. We obtain
\begin{eqnarray}
\dot{\rho}^{I}_{I} &=& - 4 \Gamma_0 \rho^I_I \label{ME1}\\
\dot{\rho}^{I}_- &=& ( -i \omega_o - 3 \Gamma_0 + i \sigma +
\Gamma_r)
\rho^{I}_- \\
\dot{\rho}^{I}_+ &=& ( -i \omega_o - 3 \Gamma_0 - i \sigma -
\Gamma_r)
\rho^{I}_+ \\
\dot{\rho}^{I}_{O} &=& (-2 i \omega_o - 2 \Gamma_0) \rho^{I}_{O} \\
\dot{\rho}^-_{O} &=& = (-i \omega_o - \Gamma_0 +i \sigma + \Gamma_r)
\rho^-_{O} + (\Gamma_0 + \Gamma_r) \alpha_1(t) \rho^{I}_+ +
(\Gamma_0 -
\Gamma_r) \alpha_2(t) \rho^{I}_- \\
\dot{\rho}^+_{O} &=& = (-i \omega_o - \Gamma_0 +i \sigma + \Gamma_r)
\rho^+_{O} + (\Gamma_0 + \Gamma_r) \alpha_3(t) \rho^{I}_+ +
(\Gamma_0 -
\Gamma_r) \alpha_4(t) \rho^{I}_- \\
\dot{\rho}^+_{+} &=& -2 (\Gamma_0 + \Gamma_r) \rho^+_+ + (\Gamma_0 +
\Gamma_r) \alpha_5(t) \rho^{I}_{I} \\
\dot{\rho}^-_- &=& -2 (\Gamma_0 - \Gamma_r) \rho^+_+ + (\Gamma_0 -
\Gamma_r) \alpha_6(t) \rho^{I}_{I} \\
\dot{\rho}^+_- &=& 2(i\sigma - \Gamma_0) \rho^+_-. \label{MEf}
\end{eqnarray}
Explicit expressions for the functions of time $\alpha_i(t), i = 1,
\ldots, 6$ appearing in Eqs. (\ref{ME1}--\ref{MEf}) are given in the
Appendix B.

We see that the evolution equation for the reduced density matrix of
the two qubits, while it is local-in-time, it does not have
constant-in-time coefficients. Hence, it does not correspond to a
Markov master equation of the Lindblad type. Again, we note that the
non-Markovian behavior is solely found in the off-diagonal terms of
the evolution law and that the diagonal ones involve constant
coefficients.

To facilitate comparison with the expressions obtained from the
Born-Markov approximations we cast equations (\ref{ME1}--\ref{MEf})
into an operator form.
\begin{eqnarray}
\dot{\hat{\rho}} &=& - i [\hat{H}_0 + \hat{H}_i, \hat{\rho}] +
\sum_{i,j=1}^2 \Gamma_{ij} (\hat{S}_+^{(i)} \hat{S}_-^{(j)}
\hat{\rho} + \hat{\rho} \hat{S}_+^{(i)} \hat{S}_-^{(j)} - 2
\hat{S}_-^{(i)} \hat{\rho} \hat{S}_+^{(j)}) \nonumber \\
&+& (\Gamma_0 + \Gamma_r) {\bf F_t}[\hat{\rho}] + (\Gamma_0 -
\Gamma_r) {\bf G_t}[\hat{\rho}]. \label{Lindbladlike}
\end{eqnarray}

The first term on the right-hand-side of Eq. (\ref{Lindbladlike})
corresponds to the usual Hamiltonian evolution: the total
Hamiltonian is a sum of the free Hamiltonian $\hat{H}_0$ and of a
dipole interaction Hamiltonian
\begin{eqnarray}
\hat{H}_i = - \sigma (\hat{S}^- \otimes \hat{S}_+ + \hat{S}^+
\otimes \hat{S}^-).
\end{eqnarray}

The second term in the right-hand-side of Eq. (\ref{Lindbladlike})
is the usual Lindblad term (see e.g., \cite{FicTan}), where  we
defined $\Gamma_{11} = \Gamma_{22} = \Gamma_0$ and $\Gamma_{12} =
\Gamma_{21} = \Gamma_r$.

The last two terms contain effects that pertain to the off-diagonal
terms of the reduced density matrix propagator and they are
non-Markovian: ${\bf F_t}$ and ${\bf G_t}$ are trace-preserving
linear operators on the space of density matrices and their explicit
form is given in Appendix B. Assuming that the Markovian regime
corresponds to the vanishing of ${\bf F_t}$ and ${\bf G_t}$, we find
from Eqs. (\ref{FF}--\ref{GG}) that in this regime the functions
$\alpha_i(t)$ in Eqs. (\ref{ME1}--\ref{MEf}) should reduce to the
following constants
\begin{eqnarray}
\alpha_1 = \alpha_4 = 0; \hspace{2cm}  \alpha_3 = \alpha_5 =
\alpha_6 = 2; \hspace{2cm} \alpha_2 = -2. \label{coefficients}.
\end{eqnarray}

In Sec. V, we shall discuss in more detail the physical origin of
non-Markovian behavior in this two-qubit system.

\section{Disentanglement of two qubits}

In this section, we employ the results obtained above to study the
evolution of the two qubits initially in an entangled state. We
shall focus on the process of disentanglement induced by their
interaction with the field.

\subsection{Class A states: Initial superposition of $|00 \rangle$ and $|11\rangle$}

We first examine the class of initial states we call Class A of the
following type
\begin{eqnarray}
| \psi_o \rangle = \sqrt{1-p} |00\rangle + \sqrt{p} |11 \rangle,
\label{gggg}
\end{eqnarray}
where $0 \leq p \leq 1$. Recall our definition $|I \rangle = |11
\rangle$ and  $|O \rangle = |00 \rangle$. From Eqs. (\ref{1111} -
\ref{0000}) we obtain
\begin{eqnarray}
\hat{\rho}(t) = p^2 e^{-4 \g_0 t} |I \rangle \langle I| + e^{-2 \g_0
t} \sqrt{p(1-p)} (  e^{2 i \omega_o t} |I\rangle \langle O| +
 e^{-2 i \omega_0 t}  \, |O \rangle \langle I|) \nonumber
\\ + [ \kappa_1(t)
- \kappa_2(t)] |- \rangle \langle -| + [\kappa_1 (t) + \kappa_2(t)]
|+ \rangle \langle +| + [1 - p^2 e^{-4 \g_0 t} - 2 \kappa_1(t)] |O
\rangle \langle O|,
\end{eqnarray}
where the functions $\kappa_1(t)$ and $\kappa_2(t)$ are given by
Eqs. (\ref{kappa1}--\ref{kappa2}).

These results are quite different from those reported in Ref.
\cite{FicTan06}, which were obtained under the Born-Markov
approximation. While the $|I \rangle \langle I|$ and $|I\rangle
\langle O|$ terms are essentially the same, the $|- \rangle \langle
-|$ and $|+ \rangle \langle +|$ ones are not, as they involve
non-diagonal elements of the density matrix propagator. For
comparison, we reproduce here the explicit form of these matrix
elements in our calculation
\begin{eqnarray}
\rho^+_+ &=& p \frac{\Gamma_0 + \Gamma_r}{\Gamma_0 - \Gamma_r} e^{-2
\Gamma_0t} (e^{-\Gamma_0 t} - e^{- \Gamma_rt})^2 \\
\rho^-_- &=& p e^{-2 \Gamma_0t} (e^{-\Gamma_0 t} - e^{-
\Gamma_rt})^2,
\end{eqnarray}
and in that of Ref. \cite{FicTan06}
 (translated into our notation):
\begin{eqnarray}
\rho^+_+ (t) &=& p \frac{\Gamma_0 + \Gamma_r}{\Gamma_0 - \Gamma_r}
e^{-2 \Gamma_0 t} ( e^{-2 \Gamma_rt} - e^{- 2 \Gamma_0 t}) \label{FTt} \\
\rho^-_-(t) &=&  p \frac{\Gamma_0 - \Gamma_r}{\Gamma_0 + \Gamma_r}
e^{-2 \Gamma_0 t} ( e^{2 \Gamma_rt} - e^{- 2 \Gamma_0 t}).
\label{FTs}
\end{eqnarray}
For large values of $r$, $\Gamma_r << \g_0$ and the expressions
above coincide. However, for $\w_0 r << 1$ their difference is
substantial. In this regime, $\g_0 \simeq \Gamma_r$ and at times
$\g_0 t \sim 1$ we obtain $(\g_0 - \Gamma_r)t <<1$. According to the
Markovian results of \cite{FicTan06}, in this regime the $|+ \rangle
\langle +|$ term is of order $O(\lambda^0)$ and hence comparable in
size to the other terms appearing in the evolution of the density
matrix. However, according to our results, which are based on the
full non-Markovian dynamics, the $|+ \rangle \langle +|$ term is of
order $\frac{\g_0 - \Gamma_r}{\g_0}$ and hence much smaller.

In general, for $\w_0 r << 1$, we find that  the $|- \rangle \langle
-|$ and $|+ \rangle \langle +|$ terms contribute little to the
evolution of the reduced density matrix and they can be ignored.
Since these terms are responsible for the sudden death and
subsequent revival of entanglement studied in \cite{FicTan06}, we
conclude that these effects are absent in this regime. Indeed, this
can be verified by the study of concurrence as a function of time as
appearing in Figs. 1 and 2. For large inter-qubit separations and
for specific initial states ($ p > 0.5$) there is sudden death of
entanglement--but no subsequent revivals.

\begin{figure} \label{ghzconc}
\centerline{\scalebox{.8}{\includegraphics{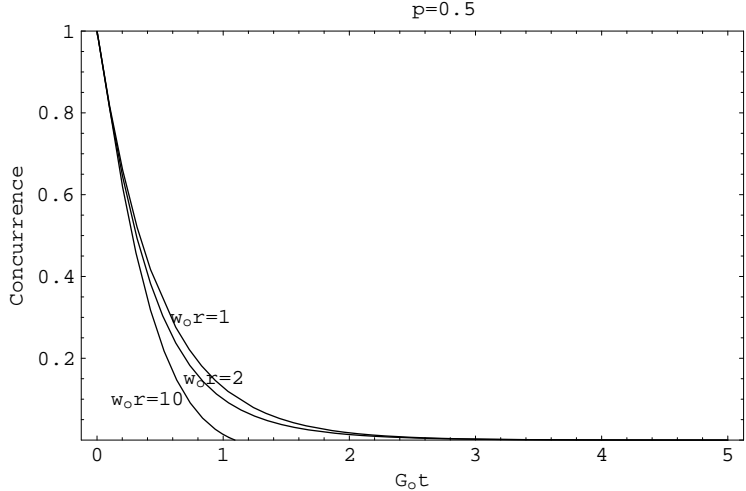}}}
\caption{A plot of the concurrence as a function of time for initial
Class A state (\ref{gggg}) with $p = 0.5$  for different values of
$\w_o r$.}
\end{figure}

\begin{figure} \label{ghzconc2}
\centerline{\scalebox{.8}{\includegraphics{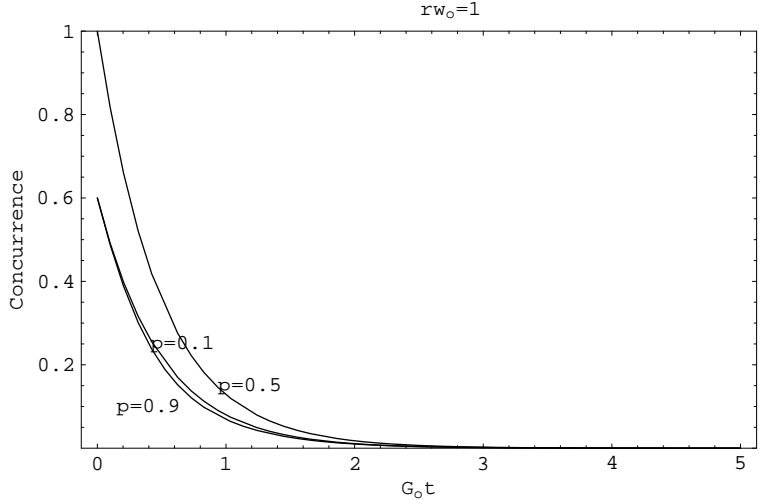}}}
\caption{A plot of the concurrence as a function of time for initial
Class A state (\ref{gggg})  for different values of $p$ and fixed
inter-qubit distance ($\w_o r = 1$).}
\end{figure}

\subsection{Initial antisymmetric $|- \rangle$  Bell state}

We next consider the case of an initial antisymmetric $|- \rangle$
state for the  two qubits. We find
\begin{eqnarray}
\hat{\rho}(t) = e^{ -2 [\g_0 - \Gamma_r]t} |- \rangle \langle -| +
\left(1 -e^{ -2 [\g_0 - \Gamma_r]t}\right)|O \rangle \langle O |
\end{eqnarray}

We see that the decay rate is $\g_0 - \Gamma_r$. The effect of
photon exchange between the qubits essentially slows down the
overall emission of photons from the two-qubit system (sub-radiant
behavior). As the qubits are brought closer, the decay is slower. At
the limit  $r \rightarrow 0$, there is no decay.

The results agree qualitatively with those obtained in Ref.
\cite{FicTan} through the Born-Markov approximation. Fig. 3 shows
the time evolution of concurrence for an initial antisymmetric $|-
\rangle$ state and for different inter-qubit distances.

\begin{figure} \label{singlet conc}
\centerline{\scalebox{1.0}{\includegraphics{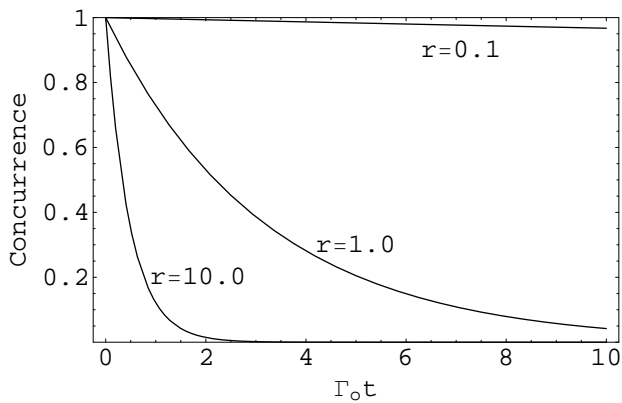}}}
\caption{A plot of the concurrence as a function of time for the
initial antisymmetric $|- \rangle$ Bell state and for different
values of the inter-qubit distance $r$ (in units of $\w_o^{-1}$).
The decay of concurrence proceeds more slowly when the qubits are
closer together.}
\end{figure}

\subsection{Initial symmetric $|+ \rangle$  Bell state}

For an initial symmetric $|+ \rangle$  Bell state the reduced
density matrix of the two qubits is

\begin{eqnarray}
\hat{\rho}(t) = e^{ -2 [\g_0 + \Gamma_r]t} |+ \rangle \langle +| +
\left(1 -e^{ -2 [\g_0 + \Gamma_r]t}\right) |O \rangle \langle O|.
\end{eqnarray}

Here we have a super-radiant decay with rate given by $\g_0 +
\Gamma_r$. Note that for both the antisymmetric $|- \rangle $  and
symmetric $|+ \rangle $  states, a resonance appears when $r = n \pi
\omega_o^{-1}$, and the decay becomes radiant. Fig. 4 shows the
behavior of concurrence, in qualitative agreement with the
corresponding results of Ref. \cite{FicTan}.

\begin{figure} \label{triplet conc}
\centerline{\scalebox{.8}{\includegraphics{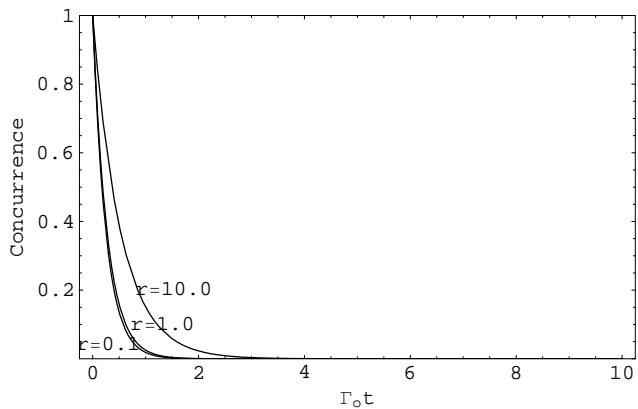}}}
\caption{Plot of the concurrence as a function of time for initial
symmetric $|+ \rangle$ Bell state and for different values of the
inter-qubit distance $r$ (in units of $\w_o^{-1}$).}
\end{figure}

\section{Decoherence of qubits}

In this section we study the evolution of factorized initial states,
in order to understand how the decoherence of one qubit is affected
by the presence of another.

\subsection{One qubit in vacuum state}

We assume that the first qubit is prepared initially in a
superposition of the $|0 \rangle$ and $|1 \rangle$ states, and that
the second qubit lies on the ground state. Therefore, the initial
state is of the form
\begin{eqnarray}
\left(\sqrt{p} |1 \rangle + \sqrt{1-p} |0 \rangle \right) \otimes |0
\rangle =  \sqrt{p} |10 \rangle + \sqrt{1-p} |00 \rangle.
\end{eqnarray}

From Eqs. ( \ref{1111}--\ref{0000}), we obtain the density matrix of
the combined qubit system
\begin{eqnarray}
\hat{\rho}(t) = p \left( |v_+|^2 |10\rangle \langle 10| + |v_-|^2
|01 \rangle \langle 01 | + v_-^* v_+ |01 \rangle \langle 10 | +
v_- v_+^* |10 \rangle \langle 01| \right) \nonumber \\
+ \sqrt{p(1-p)} \left( v_+^* |10 \rangle \langle 00| + v_-^* |01
\rangle \langle 00| + v_+ |00 \rangle \langle 10| + v_- |00 \rangle
\langle 01| \right) \nonumber \\
+ \left(1 - p (|v_+|^2 + |v_-|^2) \right) |00 \rangle \langle 00|,
\end{eqnarray}
where the functions $v_{\pm}(t)$ are given by Eq. (\ref{v+-}).

The two qubits become entangled through their interaction via the EM
field bath. To study the  decoherence in the first qubit, we trace
out the degrees of freedom of the second one, thus constructing the
reduced density matrix $\hat{\tilde{\rho}}_1$
\begin{eqnarray}
\hat{\tilde{\rho}}_1(t) = p |v_+|^2 |1 \rangle \langle 1| +
\sqrt{p(1-p} \left( v_+ |0\rangle \langle 1| + v_+^* |1 \rangle 0|
\right) + \left( 1 - p|v_+|^2 \right) |0 \rangle \langle 0|.
\end{eqnarray}

At the limit of large interqubit distances, $\Gamma_r = 0 =
\sigma(r)$, whence $v_+ \simeq e^{- i \w_o t - \g_0 t}$, the
off-diagonal elements decay within a characteristic time-scale of
order $\g_0^{-1}$. These results coincide with those for the single
qubit case--see Refs. \cite{AH, SADH}. However, in the  regime $\w_o
r << 1$, the results are substantially different. The entanglement
with the second qubit leads to a departure from pure exponential
decay. In this regime, $\Gamma_r \simeq \g_0$. This implies for
times longer than $\g_0^{-1}$ a substantial fraction of the
off-diagonal elements persists. This decays eventually to zero
within a time-scale of order $[\g_0 - \Gamma_r]^{-1} >> \g_0^{-1}$.
Hence, the qubit preserves its coherence longer. (At the limit $r
\rightarrow 0$ there is no decoherence.)

The reduced density matrix of the second qubit is
\begin{eqnarray}
\hat{\tilde{\rho}}_2(t) = p |v_-|^2 |1 \rangle \langle 1| +
\sqrt{p(1-p)} \left( v_- |0\rangle \langle 1| + v_-^* |1 \rangle 0|
\right) + \left( 1 - p |v_-|^2 \right) |0 \rangle \langle 0|.
\end{eqnarray}

Note that at the limit of small inter-qubit distances the asymptotic
behavior ( for $\g_0 t >> 1$) of $\hat{\tilde{\rho}}_1(t)$ is
identical to that of $\hat{\tilde{\rho}}_2(t)$. The second qubit
(even though initially on its ground state) develops a persistent
quantum coherence as a result of the interaction with the first one.

\subsection{One qubit in excited state}

We also consider the case of a factorized initial state, in which
the second qubit is excited
\begin{eqnarray}
\left(\sqrt{p} |1 \rangle + \sqrt{1-p} |0 \rangle \right) \otimes |1
\rangle =  \sqrt{p} |I \rangle + \sqrt{1-p} |01 \rangle.
\end{eqnarray}
This system behaves differently from that of Sec. IV.A. The matrix
elements of the reduced density matrix read
\begin{eqnarray}
\rho^{11}_{11} &=& p e^{ - 4 \g_0 t} \\
\rho^{11}_{01} &=& \sqrt{p(1-p)} e^{ - 2 i \w_o t - 2 \g_0 t} v_-^* \\
\rho^{01}_{00} &=& \sqrt{p(1-p)} \mu_1(t) \\
\rho^{10}_{00} &=& \sqrt{p(1-p)} \mu_2^*(t)\\
\rho^{01}_{01} &=& (1-p) |v_+|^2 + p \kappa_1 \\
\rho^{01}_{10} &=& (1-p) v_+ v^*_- + p \kappa_2,
\end{eqnarray}

where the functions $\mu_{1,2}$ are given by Eqs.
(\ref{mu1}--\ref{munu}), the functions $\kappa_{1,2}$ by Eqs.
(\ref{kappa1}) and (\ref{kappa2}) and the functions $v_{\pm}$ by Eq.
(\ref{v+-}).

The reduced density matrix of the first qubit is
\begin{eqnarray}
\hat{\tilde{\rho}}_1(t) = \left((1-p)|v_+|^2 + p \kappa_1 + p e^{- 4
\g_0t} \right) |1 \rangle \langle 1| + \left(1 - (1-p)|v_+|^2 - p
\kappa_1 - p e^{- 4 \g_0t} \right) |0 \rangle \langle 0| \nonumber\\
+ \sqrt{p(1-p)} \left( (\mu_2^* + e^{ -2 i \w_o t - 2 \g_0 t} v_+^*)
|0 \rangle \langle 1| + (\mu_2 + e^{ 2 i \w_o t - 2 \g_0 t} v_+^*)
|1 \rangle \langle 0| \right).
\end{eqnarray}

At times $t >> \g_0^{-1}$, the decay of  the off-diagonal elements
at is dominated by the function $\mu_2$, which then reads
\begin{eqnarray}
\mu_2(t) \simeq \frac{\Gamma_r e^{- i (\w_o - \sigma) t - \Gamma_0
t} }{\Gamma_0 + \frac{ 2 \sin \w_o r}{\w_o r} \Gamma_r - i \sigma (
1 + 2 \frac{\sin \w_o r}{\w_o r})}  ( e^{ - \Gamma_r t} - e^{-
\Gamma_0t} ).
\end{eqnarray}

 In the limit of large inter-qubit distance $r$, the
off-diagonal element falls like $r^{ - 2 \g_0t}$, while in the small
$r$ limit it falls like $e^{ - \g_0 t}$. Comparing with the case of
Sec. IV.A, we see that the initial excitation of the second qubit
leads to a lesser degree of entanglement of the total system, as it
cannot absorb any virtual photons emitted from the first one. For
this reason, the asymptotic decoherence rate does not vary
significantly with the interqubit distance.

\section{Discussion}

In this section, we first summarize our results for the evolution of
various initial states, we then discuss the origin of the
non-Markovian behavior and finally, a possible limitation of our
results that is due to the restricted domain of validity of the
Rotating-Wave approximation.

\subsection{Description of the results}

For an initial Class A state  (\ref{gggg}),  the $|I \rangle$
component decays to the vacuum, but it also evolves into a linear
combination of antisymmetric $|- \rangle $  and symmetric $|+
\rangle $  Bell states. However, if the qubits are close together
the evolution to Bell states is suppressed. This behavior is
qualitatively different from that of Ref. \cite{FicTan06}, which was
obtained through the Born-Markov approximation. The corresponding
terms differ by many orders of magnitude at the physically relevant
time-scales. As a consequence, we find that there is neither sudden
death nor revival of entanglement in this regime.

In retrospect, this difference should not be considered surprising.
The Born-Markov method involves two approximations: i) that the
back-action of the field on the atoms is negligible and ii) that all
memory effects in the system are insignificant. When the qubits are
found within a distance much smaller than the one corresponding to
their characteristic wavelength, it is not possible to ignore
back-action. The virtual photons  exchanged by the qubits (at a rate
given by $\Gamma_r$) substantially alter the state of the
electromagnetic field.

On the other hand, the effect of virtual photons exchange between
qubits drops off quickly at large separations $r$  -- the two qubits
decay almost independently one of the other. Hence, the Born-Markov
approximation -- reliable for the case of two separate qubits each
interacting with an individual field -- also gives reasonable
results for the two qubits interacting with a common field. In this
regime sudden death is possible, but not revival of entanglement. In
this sense, our results effectively reduces to those of Ref.
\cite{YuEbePRL}: when the distance between the qubits is much larger
than any characteristic correlation length scale of the system it
looks as though the two qubits are found in different reservoirs.

Therefore, our results for initial states of Bell type are
qualitatively similar to those obtained in Ref. \cite{FicTan}
through the Born-Markov approximation. The symmetric $|+ \rangle $
state decays super-radiantly and the antisymmetric $|- \rangle $ one
sub-radiantly. For small values of $\w_o r$, the antisymmetric $|-
\rangle $ state decays very slowly and entanglement is preserved.

Concerning decoherence, when the inter-qubit separation is large and
the second qubit lies on its ground state, our two qubit calculation
reproduces previous results on the single qubit case \cite{AH,
SADH}. However, if the qubits are close together the coherence is
preserved longer. These results seem to suggest that in a many-qubit
system, the inter-qubit quantum coherence can be sustained for times
larger than the decoherence time of the single qubit case. This may
suggest some physical mechanisms to resist decoherence in
multi-qubit realistic systems.

\subsection{The origin and significance of the non-Markovian
behavior}

With the exact solutions we have obtained (under weak coupling but
no Born or Markov approximation) for the two qubit - EMF system we
want to elaborate on the origin and significance of the
non-Markovian behavior started in the Introduction. In the evolution
equations (\ref{dmp1}--\ref{dmpf}) we note that the diagonal terms
of the reduced density matrix propagator all decay exponentially.
Their decay rate is therefore constant. It is well known that this
feature is a sign of Markovian behavior. In fact, it characterizes
the domain of validity of Fermi's golden rule: one could  obtain the
decay rates by direct application of this rule. Hence, as far as
this part of the evolution is concerned, our results are fully
compatible with the Markovian predictions.

However, the behavior of the non-diagonal terms in the reduced
density matrix propagator is different. A look at Eqs.
(\ref{dmp1}-\ref{dmpf}) will convince the reader that the only
non-zero such terms are ones that describe the effect of successive
decays, for example the $|11 \rangle $ state first decaying into $|-
\rangle$ and then $|- \rangle $ decaying into the ground state $|00
\rangle$. Hence, the $\rho^-_-(t)$ term consists of one component
that contains the remaining of the $|- \rangle \langle -|$ part of
the initial state and another component that incorporates the decay
of the $|11 \rangle \langle 11|$ part of the initial state towards
the state $|- \rangle$. In our calculation, the latter term is
encoded into the functions $\kappa_{1,2}(t)$, which are obtained by
squaring the amplitudes $\nu_{\bf k}(t)$ as in Eqs.
(\ref{defk1}--\ref{defk2}). At second loop order, these amplitudes
are obtained from the summation of two Feynman diagrams --see Eq.
(\ref{nuF}). The structure of the poles in Eq. (\ref{nuF}) reveals
that the first Feynman diagram describes the decays of the $|11
\rangle$ state, while the second one corresponds to processes
involving the $|01 \rangle$ and $|10 \rangle$ states. When we
construct the evolution functions $\kappa_{1,2}(t)$, we obtain terms
that are both diagonal and off-diagonal with respect to the two
types of process. It is the presence of the off-diagonal terms that
is primarily (but not solely) responsible for the deviation of our
results from the Markovian prediction. In the Markov approximation,
the corresponding term involves summation (subtraction) of
probabilities rather than amplitudes.

To justify the last statement, we note from Eqs.
(\ref{1111}--\ref{0000}) that the evolution decouples the diagonal
from the off-diagonal elements of the density matrix (the Markovian
equations also have this property). Hence, the probabilities $p_a(t)
= \rho^a_a(t), a \in \{ I, +, -, O\}$ evolve autonomously.
Time--homogeneity (i.e. Lindblad time evolution) implies that their
evolution can be given by a transfer matrix
\begin{eqnarray}
\dot{p}_a(t) = \sum_b T_a^b p_b(t). \label{transf}
\end{eqnarray}
Here $T^a_b$ is the {\em constant} decay rate for the process $ b
\rightarrow a$. Noting that Eqs. (\ref{1111}--\ref{0000}) contain no
transitions between $+$ and $-$ and no transitions from $-$ to $I$,
Eq. (\ref{transf}) yields
\begin{eqnarray}
\dot{p}_-(t) = - 2(\Gamma_0 - \Gamma_r) p_-(t) + w p_{I}(t),
\label{ps}
\end{eqnarray}
where $w = T^-_{I}$. Since $p_{I}(t)$ is determined  by Eq.
(\ref{ME1}) as $p_{I}(0) e^{-4 \g_0t}$, we obtain for the initial
condition $p_-(0) = 0$,
\begin{eqnarray}
p_-(t) \sim p_{I}(0) (e^{-2 (\Gamma_0 - \Gamma_r)t} - e^{-4 \Gamma_0
t}), \label{ccc}
\end{eqnarray}
in full agreement with Eq. (\ref{FTs}) obtained from the Lindblad
master equation.

The derivation of Eq. (\ref{ccc}) provides an example of a more
general fact: the Markovian assumption forces the off-diagonal terms
of the density matrix propagator (in this case the one mapping
$\rho^{I}_{I}(0)$ to $\rho^-_-(t)$) to be subsumed by the diagonal
ones. The off-diagonal elements can have no independent dynamics of
their own, unlike what would be the case if they were derived from a
full calculation. One should also note that  the diagonal terms of
the propagator are obtained from the first order perturbation
theory. Since they determine the off-diagonal terms within the
Markov approximation, the latter only contains the information
obtained from first-order perturbation theory. However, in our
calculation the off-diagonal terms involve second--order effects and
hence, reveals dynamical correlations that are inaccessible within
the context of the Born-Markov approximation.

For the reasons above, the matrix elements (\ref{FTt}--\ref{FTs})
given by \cite{FicTan06} obtained under the Markovian
approximation--carry the characteristic superradiant and subradiant
behavior of the decay of the $|+ \rangle$ and $|- \rangle$ states
respectively, a property that does not arise from our calculation.

Moreover, we note that Eq. (\ref{ps}) obtained by making the Markov
approximation  is related to arguments from {\em classical
probability}, i.e. it involves addition of the probabilities
associated to different processes \footnote{To be precise, in an
exponential decay, it is possible to define a probability density
for the occupation number of any state as a function of time. This
is in general not possible in quantum theory. Eq. (\ref{ccc}) is
then essentially the Kolmogorov additivity condition for these
probabilities. }. On the other hand, a proper quantum mechanical
calculation involves the addition of amplitudes--e.g. the ones
appearing in the definition of the functions $\kappa_{1,2}(t)$-- and
as such it must also contain `interference' terms.  In effect, the
Markovian approximation introduces by hand a degree of partial
`decoherence' (or classicality)  and we think that this explains why
in general it predicts a faster classicalization of the system than
the full analysis does.

We believe that the feature discussed above is generic. In the
single qubit system, it was not present because its structure was
too simple. There could be no intermediate decays. However, this
effect should in principle be present in any system that contains
 intermediate states. The Markovian approximation would then be
valid only if specific conditions hold that render the
`interference' terms negligible--for example if there exists a sharp
separation of the relevant timescales.

To summarize, the Markov approximation essentially ignores
interference terms that are relevant to processes that involve
successive decays. These processes appear through off-diagonal terms
in the reduced density matrix propagator. The Markov approximation
misrepresents the intrinsic dynamics for these terms and ties them
--by forcing additivity of probabilities--to the evolution of the
diagonal ones. As a results, the off-diagonal terms are subsumed by
the diagonal ones.

\subsection{The use of the rotating wave approximation}
Finally, we add a few words on the accuracy of our model for the
interaction between the 2LA's and the EMF. The Hamiltonian
(\ref{Hint}) is obtained from the study of the interaction of the
atomic degrees of freedom with the electromagnetic field. Its
derivation involves the dipole and the rotating wave (RW)
approximations--see Appendix A in Ref. \cite{AH}. The RW
approximation consists in ignoring rapidly oscillating terms in the
interaction-picture Hamiltonian and keeping only the ones that
correspond to resonant coupling. (One ignores processes during which
a photon is emitted {\em and} the atom becomes excited.) The terms
that are dropped out in the RW approximation oscillate with a
frequency of order $\w_o$.

For a single qubit system the RW approximation is self-consistent.
However, in the two-qubit system we keep terms in the Hamiltonian
that vary in space as $e^{i {\bf k} \cdot {\bf r}}$. For the RW
approximation to be consistent, one has to assume that ${\bf k}
\cdot {\bf r} << \w_o t$. Since ${\bf k}$ is peaked around the
resonance frequency, this condition is equivalent to $ r << t$.
Since the physically interesting time-scale for the study of
disentanglement and decoherence corresponds to $\g_0^{-1} \sim
[\lambda^{2} \w_0]^{-1}$, we expect the RW approximation to be
reliable in this context, as long as $\lambda^2 \w_o r << 1$. This
is sufficient for realistic situations, in which $r$ is bounded by
the size of the cavity and $\lambda^2 << 1$. However, the condition
above does not hold at the formal limit $r \rightarrow \infty$.  In
this regime the RW approximation may break down. Indeed, in section
IV.B the reduced density matrix for the single qubit in the limit $r
\rightarrow \infty$ does not coincide with the corresponding
expression obtained in the study of the single qubit. The presence
of an {\em excited}  second qubit, even if it is situated far away,
affects the time evolution significantly. This effect is also
present in the Born-Markov approximation. This is arguably an
unphysical behavior, and we believe that it arises as an artefact of
the RW approximation. \\

\noindent{\bf Acknowledgement} It is a pleasure for one of us (BLH)
to hear the lectures of Dr. Ting Yu and the seminar of Professor Joe
Eberly explicating their work on disentanglement of two qubits. We
also thank the referee for urging us to derive the master equation
so the non-Markovian features can be demonstrated more explicitly.
This work is supported by grants from the NSF-ITR program
(PHY-0426696), NIST and NSA-LPS. CA was supported by a Pythagoras II
grant (EPEAEK).

\newpage
\begin{appendix}
\section{Perturbative resummation}

We notice that the action of the operator $[(E-H_0)^{-1}H_I]^2$ on
any linear combination of the vectors $| O;0,1 \rangle $ and $| O;
1, 0 \rangle$ yields another linear combination of these vectors. If
we denote by $p_n $ and $q_n$ the coefficients of $| O;0,1 \rangle $
and $| O; 1,0 \rangle$ respectively after the n-th action of
$[(E-H_0)^{-1}H_I]^2$ (2n-th order of perturbation theory) we obtain
\begin{eqnarray}
\left( \begin{array}{l}  p_n \\ q_n \end{array} \right) = \frac{1}{E
- \Omega}\left(
\begin{array}{ll} \alpha(E) & \beta^*(E,r) \\
\beta(E,r) & \alpha(E) \end{array} \right) \left( \begin{array}{l}
p_{n-1} \\ q_{n-1} \end{array} \right) = : A \left(
\begin{array}{l} p_{n-1} \\ q_{n-1} \end{array} \right)
\end{eqnarray}
Consequently,
\begin{eqnarray}
\left( \begin{array}{l}  p_n \\ q_n \end{array} \right) = A^n \left(
\begin{array}{l}  p_0 \\ q_0 \end{array} \right),
\end{eqnarray}
and summing over all $n$ we obtain
\begin{eqnarray}
\sum_{n=0}^{\infty} \left( \begin{array}{l}  p_n \\ q_n
\end{array} \right) = \left( \sum_{n=0}^{\infty} A^n \right)
\left( \begin{array}{l}  p_0 \\ q_0 \end{array} \right) = (1
-A)^{-1} \left( \begin{array}{l}  p_0 \\ q_0 \end{array} \right)
\end{eqnarray}

To compute $\langle z; 0,1|(E-H)^{-1}|O;01 \rangle$ and $\langle z;
1,0|(E-H)^{-1}|O;01 \rangle$ according to the perturbation series we
have $p_0 = (E-\w_o)^{-1}$, $q_0 = 0$. The results in the main text
then follow.

The resummation for $\langle z; 0,0|(E-H)^{-1}|O;0,1 \rangle$
proceeds from the fact that the  terms in the $2n+1$ order of the
perturbative expansion are of the form
\begin{eqnarray}
\sum_{\bf k} \frac{g_{\bf k} \zeta^n_{\bf k} e^{i {\bf k} \cdot {\bf
r}/2} }{E - \omega_{\bf k}} b^{\dagger}_k | O, 0 , 0 \rangle
\end{eqnarray}
We then note that
\begin{eqnarray}
\zeta_{\bf k}^n = [\frac{ \alpha(E) + \beta(E,r)e^{i {\bf k} \cdot
{\bf r}} }{E - \w_o}] \zeta_{\bf k}^{n-1},
\end{eqnarray}
where $\zeta^0_{\bf k} = (E- \omega_o)^{-1}$. This geometric series
can be summed to give the result quoted in the main text.

A similar summation is achieved for the terms $\langle z;
0,1|(E-H)^{-1}|O; 1, 1 \rangle $ and $\langle z; 1, 0|(E-H)^{-1}|O;
1, 1 \rangle $.  In the $2n$-th order of perturbation theory the
action of the series on $O; 1, 1 \rangle$ yields
\begin{eqnarray}
\sum_{\bf k} \frac{g_{\bf k} }{(E - 2 \Omega) (E - \Omega -
\omega_{\bf k})} \left( e^{-i {\bf k}\cdot {\bf r}/2} \gamma^n_{\bf
k} |O; 0, 1 \rangle + e^{i {\bf k}\cdot {\bf r}/2} \delta^n_{\bf k}
|O; 1, 0 \rangle \right)
\end{eqnarray}

\section{The time-dependent terms of the master equation}

The time-dependent coefficients $\alpha_i, i = 1, \ldots 6$
appearing in the master equation (\ref{ME1}--\ref{MEf}) are defined
as
\begin{eqnarray}
\alpha_1(t) &=& i \frac{d}{d t}\left(\nu(t) e^{i \omega_ot +
\Gamma_0t - i \sigma t - \Gamma_rt}\right) e^{2 \Gamma_0 t + 2 i
\sigma t + 2 \Gamma_rt} \label{alpha1} \\
\alpha_2(t) &=& \frac{d}{d t} \left( \mu(t) e^{i\omega_0t + \Gamma_0
t- i \sigma t- \Gamma_r t} \right) e^{2 \Gamma_0t}
\\
\alpha_3(t) &=& \frac{d}{d t} \left( \mu(t) e^{i\omega_0t + \Gamma_0
t + i \sigma t + \Gamma_r t} \right) e^{2
\Gamma_0t} \\
\alpha_4(t) &=& i \frac{d}{d t}\left(\nu(t) e^{i \omega_ot +
\Gamma_0t + i \sigma t + \Gamma_rt}\right) e^{2 \Gamma_0
t - 2 i \sigma t - 2 \Gamma_rt}\\
\alpha_5(t) &=& \frac{d}{d t}\left( \kappa(t)
e^{2\Gamma_0t + 2 \Gamma_rt}\right) e^{2 \Gamma_0t- 2 \Gamma_r t} \\
\alpha_6(t) &=& \frac{d}{d t}\left( \kappa(t) e^{2\Gamma_0t - 2
\Gamma_rt}\right) e^{2 \Gamma_0t+ 2 \Gamma_r t}. \label{alpha6}
\end{eqnarray}

Using the notation
\begin{eqnarray}
\hat{A}_{+I} &=& |+ \rangle \langle I|, \\
\hat{A}_{OI} &=& |O \rangle \langle I| \\
\hat{A}_{++} &=& |+ \rangle \langle +| \\
\hat{A}_{+-} &=& |+ \rangle \langle -| \\
\hat{A}_{--} &=& |- \rangle \langle -|,
\end{eqnarray}
the linear functional ${\bf F}_t$ and ${\bf G}_t$ of Eq.
(\ref{Lindbladlike}) are defined as
\begin{eqnarray}
{\bf F_t}[\hat{\rho}] &=& -[2 - \alpha_5(t)] (\hat{A}_{+I}
\hat{\rho} \hat{A}^{\dagger}_{+I} - \hat{A}_{OI} \hat{\rho}
\hat{A}_{OI}^{\dagger}) - [2 - \alpha_3(t)] \hat{A}_{OI} \hat{\rho}
\hat{A}_{++} -
[2 - \alpha_3^*(t)] \hat{A}_{++} \hat{\rho} \hat{A}_{OI}^{\dagger} \nonumber \\
 &+&
\alpha_1(t) \hat{A}_{OI} \hat{\rho} \hat{A}_{+-} + \alpha_1^*(t)
\hat{A}_{+-}^{\dagger}
 \hat{\rho} \hat{A}_{OI}^{\dagger} \label{FF}\\
{\bf G_t}[\hat{\rho}] &=& - [2 - \alpha_6(t)] (\hat{A}_{-I}
\hat{\rho} \hat{A}^{\dagger}_{-I} - \hat{A}_{OI} \hat{\rho}
\hat{A}_{OI}^{\dagger}) +[2 + \alpha_2(t)] \hat{A}_{OI} \hat{\rho}
\hat{A}_{--} +
[2 + \alpha_2^*(t)] \hat{A}_{--} \hat{\rho} \hat{A}_{OI}^{\dagger} \nonumber \\
&+& \alpha_4(t) \hat{A}_{OI} \hat{\rho} \hat{A}_{+-}^{\dagger} +
\alpha_4^*(t) \hat{A}_{+-}^{\dagger} \hat{\rho}
\hat{A}_{OI}^{\dagger} \label{GG}.
\end{eqnarray}

\section{A Grassmann path-integral derivation}

In Refs. \cite{AH,SADH} the evolution of the reduced density matrix
for a 2LA interacting with the EM field was determined with a
version of the Feynman-Vernon path integral method, using Grassmann
variables for the atomic degrees of freedom. The same method can be
applied to the present problem, even though the perturbative
approach turned out to be more convenient because the vacuum initial
state allowed for a summation of the perturbative series. However,
the path-integral method may allow for a simpler treatment of other
systems, such as an N-qubit system or the EM field at a finite
temperature. For this reason, we include a sketch of the
path-integral treatment in this Appendix, noting that it leads to
the same results as the ones obtained in Sec. II.

\subsection{Coherent state path integral} The coherent states for
the EMF (bosonic) and 2LA (qubit) states are defined respectively by
\begin{eqnarray}
\label{emcoherentstates} |z_\bk \rangle &=& \exp(z_\bk
b_\bk^\dagger)
|0_\bk \rangle \\
\label{gmncoherentstates} |\n^{(n)} \rangle &=& \exp(\n S_+^{(n)})
|0\rangle_{(n)}.
\end{eqnarray}
The states $|0_\bk \rangle$ and $|0\rangle_{(n)}$ are ground states
of the electromagnetic field's $\bk^{\mbox{th}}$ mode in vacuum and
the $n^{\mbox{th}}$ qubit in its lower state, repectively. The
bosonic coherent states are labeled by a complex number, $z_\bk$,
and the qubit coherent states are labelled by an
anti-commuting~(i.e.~Grassmannian) number, $\eta^{(n)}$. A
non-interacting eigenstate of two qubits and the electromagnetic
spectrum can be written as the direct product of coherent states,
\begin{equation}
|\{z_\bk\},\n^{(1)},\n^{(2)}  \rangle =|\{z_\bk\}\rangle \otimes
|\n^{(1)}\rangle \otimes |\n^{(2)} \rangle.
\end{equation}

In a path integral approach the transition amplitude between some
chosen initial and final states is divided into many infinitesimal
time steps. The resulting path integral corresponds to the matrix
element of the evolution operator. In the coherent-state path
integral representation the Hamiltonian is given by,
\begin{eqnarray}\label{hamiltonian2}
\frac{\langle \nb^{(1)}, \nb^{(2)}, \{\zb_{k}\} | H  |
\eta^{\prime(1)}, \eta^{\prime(2)}, \{\zp_{k}\}
\rangle}{\exp[\nb^{(1)} \eta^{\prime(1)} +\nb^{(2)} \eta^{\prime(2)}
+\sum_\bk \zb_{k} \zp_{k}]}= \hbar \omega_o \nb \np + \hbar \sum_\bk
\left[ \omega_{\bk} \zb_\bk \zp_\bk + \sum_n \left(\la_{\bk}^{(n)}
\nb^{(n)} \zp_\bk  +\lab_{\bk}^{(n)}  \zb_\bk \eta^{\prime(n)}
\right) \right].
\end{eqnarray}
Spatial dependence is carried in the coupling constants,
\begin{eqnarray}\label{spatialdependence}
\la_\bk^{(n)} = \frac{\la^{(n)}}{\omega_\bk} e^{-i\bk\cdot{\bf r}_n} \\
\lab_\bk^{(n)} = \frac{\lab^{(n)}}{\omega_\bk} e^{+i\bk\cdot{\bf
r}_n}.
\end{eqnarray}
The transition amplitude for N atoms after the j-th infinitesimal
time steps is,
\begin{equation} \label{jstepprop}
K(j\epsilon,0) = \exp\bigg\{ \sum_n \nb_{j}^{(n)} \bigg[\sum_m
\psi_j^{(nm)}\n_{0}^{(m)} +\sum_{\bk\bl}
g_{j\bk\bl}^{(n)}\z_{0\bl}\bigg] +\sum_\bk \zb_{j\bk} \bigg[\sum_\bl
f_{j\bk\bl}\z_{0\bl} +\sum_m \phi_{j\bk}^{(m)}\n_{0}^{(m)}\bigg]
\bigg\}.
\end{equation}
The functions in the path integral are determined by
finite-difference equations
\begin{equation} \label{inductive1}
\begin{array}{lll}
\psi_{j+1}^{(nm)} = (1-i \w_o \e) \psi_{j}^{(nm)} -i\e\sum_{\bk}
\la_{j\bk}^{(n)} \phi_{j\bk}^{(m)} && \psi_0^{(nm)} =\delta_{nm} \\
\phi_{j+1,\bk}^{(m)} = (1-i \w_{\bk}\e)\phi_{j\bk}^{(m)} -i\e \sum_n
\lab_{j\bk}^{(n)} \psi_{j}^{(nm)}
 && \phi_{0\bk}^{(m)} =0
\end{array}
\end{equation}
\begin{equation} \label{inductive2}
\begin{array}{lll}
g_{j+1,\bk\bl}^{(n)} = (1-i \w_o\e) g_{j\bk\bl}^{(n)} -i\e
\la_{j\bk}^{(n)} f_{j\bk\bl} && g_{0\bk\bl} = 0 \\
f_{j+1,\bk\bl} = (1-i\w_{\bk}\e) f_{j\bk\bl}  -i\e \sum_m
\lab_{j\bk}^{(m)} \sum_\bq g_{j\bq\bl}^{(m)} && f_{0\bk\bl}
=\delta_{\bk\bl}.
\end{array}
\end{equation}

\subsection{Reduced Density Matrix}
The reduced density matrix of the two 2LAs is
\begin{equation} \label{evolved1}
\rho(t) = \int
d\mu(\n_0^{(2)})d\mu(\n_0^{\prime(2)})d\mu(\n_0^{(1)})d\mu(\n_0^{\prime(1)})
\rho(0) J_R(t,0).
\end{equation}
The dynamics of this open system is described by the density matrix
propagator
\begin{equation}\label{reduced propagator}
J_R(t,0)\vert_{T=0} = \exp\bigg\{ \sum_{nm} \bigg[\nb_t^{(n)}
\psi(t)^{(nm)} \n_0^{(m)} +\nb_0^{\prime(n)} \psibp(t)^{(nm)}
\n^{\prime(m)}_t +\nb_0^{\prime(n)} \sum_\bk \phibp_{k}(t)
\phi_{k}(t) \n_0^{(m)}\bigg] \bigg\}
\end{equation}

A general initial two-qubit density matrix written in the Grassmann
representation is:
\begin{eqnarray}
\rho(0)= \left( \begin{array}{ccccccc} 1 && \nb_0^{(2)} &&
\nb_0^{(1)} && \nb_0^{(1)}\nb_0^{(2)}\end{array} \right) \left(
\begin{array}{lllllll} \rho^{00}_{00} && \rho^{00}_{01} &&
\rho^{00}_{10} && \rho^{00}_{11} \\ \\ \rho^{01}_{00} &&
\rho^{01}_{01} && \rho^{01}_{10} && \rho^{01}_{11} \\ \\
\rho^{10}_{00} && \rho^{10}_{01} && \rho^{10}_{10} && \rho^{10}_{11}
\\ \\ \rho^{11}_{00} && \rho^{11}_{01} && \rho^{11}_{10} &&
\rho^{11}_{11} \end{array} \right) \left( \begin{array}{c} 1 \\ \\
\n_0^{\prime(2)} \\ \\ \n_0^{\prime(1)}
\\ \\ \n_0^{\prime(1)}\n_0^{\prime(2)} \end{array} \right).
\end{eqnarray}

 In addition to Eqs.
Eq.(\ref{inductive1}-\ref{inductive2}), we derive finite difference
equations for the expanded functions appearing in the reduced
density matrix elements
\begin{eqnarray}
\psi_{j+1}^{(ab)}\psi_{j+1}^{(nm)} &=& (1-2i\w_o
\e)\psi_{j}^{(ab)}\psi_{j}^{(nm)} -i\e\sum_{\bk} \la_{j\bk}^{(n)}
\phi_{j\bk}^{(m)}\psi_{j}^{(ab)} -i\e\sum_{\bk} \la_{j\bk}^{(a)}
\phi_{j\bk}^{(b)}\psi_{j}^{(nm)}
\\
\psi_{j+1}^{(ab)}\phi_{j+1,\bk}^{(m)} &=&
(1-i\w_o\e-i\w_\bk)\psi_{j}^{(ab)}\phi_{j\bk}^{(m)} -i\e\sum_{n}
\lab_{j\bk}^{(n)} \psi_{j}^{(nm)}\psi_{j}^{(ab)} -i\e\sum_{\bl}
\la_{j\bl}^{(a)}
\phi_{j\bl}^{(b)}\phi_{j\bk}^{(m)} \\
\phi_{j+1,\bl}^{(b)}\phi_{j+1,\bk}^{(m)} &=&
(1-i\w_\bl\e-i\w_\bk)\phi_{j\bl}^{(b)}\phi_{j\bk}^{(m)} -i\e\sum_{n}
\lab_{j\bk}^{(n)} \psi_{j}^{(nm)}\phi_{j\bl}^{(b)} -i\e\sum_{a}
\lab_{j\bl}^{(a)} \psi_{j}^{(ab)}\phi_{j\bk}^{(m)}
\end{eqnarray}
For a two-atom system, it is convenient to represent the functions
appearing in the propagator by column vectors,
\begin{eqnarray}
\label{columnvec1} [\psi] &=& \left( \begin{array}{c} \psi^{(22)} \\
\psi^{(21)} \\ \psi^{(12)} \\ \psi^{(11)} \end{array}\right)
\;\;\;\;\;\;\;\; \label{columnvec2} [\phi_\bk] = \left(
\begin{array}{c}
\phi_\bk^{(2)} \\ \phi_\bk^{(1)} \end{array}\right) \;\;\;\;\;\;\;
\label{columnvec3} [\psi\psi] = \left( \begin{array}{c}
\psi^{(11)}\psi^{(22)} \\ \psi^{(21)}\psi^{(12)} \end{array}\right) \\
\label{columnvec4} [\psi\phi_{\bk}] &=& \left( \begin{array}{c}
\psi^{(22)}\phi_\bk^{(1)} \\ \psi^{(21)}\phi_\bk^{(2)} \\
\psi^{(12)}\phi_\bk^{(1)} \\ \psi^{(11)}\phi_\bk^{(2)}
\end{array}\right)  \;\;\;\;\;\;\;\;\; \label{columnvec5} [\phi\phi_{\bk\bl}] =
\left( \begin{array}{c} \phi_\bk^{(1)}\phi_\bl^{(2)} \\
\phi_\bk^{(2)}\phi_\bl^{(1)}
\end{array}\right).
\end{eqnarray}
At the continuous limit the finite difference equations yield a set
of differential equations for the column vectors above.
\begin{eqnarray}
\label{functional1} \frac{d}{dt} [\psi] &=& -i\w_o [\psi] -i\sum_\bk
G^{\dagger}_\bk [\phi_{\bk}] \\
\label{functional2} \frac{d}{dt} [\phi_\bk] &=& -i\w_\bk [\phi_\bk]
-iG_\bk [\psi] \\
\label{functional3} \frac{d}{dt} [\psi\psi] &=& -2i\w_o [\psi\psi]
-i\sum_\bk A_\bk [\phi\psi_{\bk}] \\
\label{functional4} \frac{d}{dt} [\phi\psi_{\bk}] &=& -i(\w_o
+\w_\bk) [\phi\psi_{\bk}] -i A^{\dagger}_\bk [\psi\psi] -i\sum_\bl
L_\bl
[\phi\phi_{\bk\bl}]\\
\label{functional5} \frac{d}{dt} [\phi\phi_{\bk\bl}] &=& -i(\w_\bl
+\w_\bk) [\phi\phi_{\bk\bl}] -i G_{\bl} [\phi\psi_{\bk}] -i D_{\bk}
[\phi\psi_{\bl}],
\end{eqnarray}
in terms of the matrices
\begin{eqnarray}
G_\bk &=& \left(\begin{array}{cccc} \lab_\bk^{(2)} & 0 &
\lab_\bk^{(1)} & 0 \\ 0 & \lab_\bk^{(2)} & 0 & \lab_\bk^{(1)}
\end{array}\right) \;\;\;\;\;\;\; A_\bk = \left(\begin{array}{cccc}
\la_\bk^{(1)} & 0 & 0 & \la_\bk^{(2)} \\ 0 & \la_\bk^{(1)} &
\la_\bk^{(2)} & 0 \end{array}\right) \\
 D_{\bk} &=&
\left(\begin{array}{cccc} 0 & \lab_\bk^{(2)} & 0 & \lab_\bk^{(1)}\\
\lab_\bk^{(2)} & 0 & \lab_\bk^{(1)} & 0
\end{array}\right) \;\;\;\;\;\;\;
L_\bl = \left(\begin{array}{cc} 0 & \la_\bl^{(2)} \\ 0 &
\la_\bl^{(2)} \\ \la_\bl^{(1)} & 0 \\ \la_\bl^{(1)} & 0
\end{array}\right)
\end{eqnarray}

The solution of Eqs. (\ref{functional1}--\ref{functional5}) allows
one to fully reconstruct the propagating amplitude and from this the
elements of the reduced density matrix for the qubits.  These
equations can be solved by implementing an approximation scheme
similar to that employed for the operator method in the main text.
We do not provide the details, but only note that the results of the
two methods coincide.

\end{appendix}

\end{document}